\documentclass[twocolumn,showpacs,showkeys,nofootinbib,superscriptaddress]{revtex4-1}

\usepackage{dcolumn}
\usepackage{bm}
\usepackage{amsmath}    % need for subequations
\usepackage{amsfonts} % note how statements can be commented out
\usepackage{amssymb}
\usepackage{mathptmx}      % use Times fonts if available on your TeX system
\usepackage{graphicx}   % for figures
\usepackage{mathtools}
\usepackage{graphics}
\usepackage{multirow}
\usepackage{bibentry}
\usepackage{braket}
\usepackage{algorithm}
\usepackage{algorithmic}
\usepackage{listings}
\usepackage{setspace}
\usepackage{slashed}
\usepackage{color}
\usepackage{soul}

\usepackage{xfrac}
\def\ep{\epsilon}

\def\la{\langle}
\def\ra{\rangle}

\def\lam{\lambda}

\def\be{\begin{equation}}
\def\ee{\end{equation}}
\def\bea{\begin{eqnarray}}
\def\eea{\end{eqnarray}}

\usepackage{hyperref}
\hypersetup{colorlinks=true,citecolor=blue,linkcolor=blue,urlcolor=blue}

\makeatletter
\providecommand \@ifxundefined [1]{%
 \@ifx{#1\undefined}
}%
\providecommand \@ifnum [1]{%
 \ifnum #1\expandafter \@firstoftwo
 \else \expandafter \@secondoftwo
 \fi
}%
\providecommand \@ifx [1]{%
 \ifx #1\expandafter \@firstoftwo
 \else \expandafter \@secondoftwo
 \fi
}%
\providecommand \href@noop [0]{\@secondoftwo}%
\providecommand \href [0]{\begingroup \@sanitize@url \@href}%
\providecommand \@href[1]{\@@startlink{#1}\@@href}%
\providecommand \@@href[1]{\endgroup#1\@@endlink}%
\providecommand \@sanitize@url [0]{\catcode `\\12\catcode `\$12\catcode
  `\&12\catcode `\#12\catcode `\^12\catcode `\_12\catcode `\%12\relax}%
\providecommand \@@startlink[1]{}%
\providecommand \@@endlink[0]{}%
\providecommand \url  [0]{\begingroup\@sanitize@url \@url }%
\providecommand \@url [1]{\endgroup\@href {#1}{\urlprefix }}%
\providecommand \urlprefix  [0]{URL }%
\providecommand \selectlanguage [0]{\@gobble}%
\providecommand \bibinfo  [0]{\@secondoftwo}%
\providecommand \bibfield  [0]{\@secondoftwo}%
\providecommand \BibitemShut  [1]{\csname bibitem#1\endcsname}%
\let\auto@bib@innerbib\@empty

\begin{document}
\title{Current-component independent transition form factors for semileptonic and rare $D\to \pi(K)$ decays in the light-front quark model}
\author{ Ho-Meoyng Choi\\
{\em Department of Physics, Teachers College, Kyungpook National University,
     Daegu, Korea 702-701}}
\email{homyoung@knu.ac.kr}

\begin{abstract}
\section*{Abstract}
We investigate the exclusive semileptonic and rare $D\to \pi(K)$ decays within the standard model together with
the light-front quark model (LFQM) constrained by the variational principle for the QCD-motivated effective Hamiltonian.
The form factors are obtained in the $q^+=0$ frame and then analytically continue to the physical timelike region.
Together with our recent analysis of the current-component independent form factors $f_\pm(q^2)$ for the semileptonic decays,
we present the current-component independent tensor form factor $f_T(q^2)$ for the rare decays to make the complete set of hadronic
matrix elements regulating the semileptonic and rare $D\to\pi(K)$ decays in our LFQM. The tensor form factor
$f_T(q^2)$ are obtained from two independent sets $(J^{+\perp}_T, J^{+-}_T)$ of the tensor current $J^{\mu\nu}_T$.
As in our recent analysis of $f_-(q^2)$, we show that $f_T(q^2)$ obtained from the two different sets of the current components
gives the identical result in the valence region of the $q^+=0$ frame without involving the explicit zero modes and the instantaneous contributions. 
The implications of the zero modes and the instantaneous contributions are also discussed in comparison between the manifestly covariant 
model and the standard LFQM.
In our numerical calculations, we obtain the $q^2$-dependent form factors $(f_\pm, f_T)$  for $D\to\pi(K)$ 
and branching ratios for the semileptonic $D\to \pi(K)\ell\nu_\ell$ ($\ell=e,\mu$) decays.
Our results show in good agreement with the available experimental data as well as other theoretical model predictions. 
\end{abstract}

%\pacs{13.40.Gp, 12.38.Lg, 13.20.He}
%\date{April 25, 2005}
\maketitle
%\section{Introduction}
\section{Introduction}
The three flavors of charged leptons, $(e, \mu, \tau)$, are the same in many respects. 
In the standard model (SM), the couplings of leptons to gauge bosons are supposed to be 
independent of lepton flavors, which is known as lepton flavor universality (LFU)~\cite{LFU1}.
The experimental tests of LFU in various semileptonic $B$ decays have been
reported~\cite{BaB12,BaB13,Belle15,Belle-sato,LHCb15} by 
measuring the ratios of branching fractions 
${\cal R}_{D^{(*)}}={\rm Br}(B\to D^{(*)}\tau\nu_\tau)/{\rm Br}(B\to D^{(*)}\ell\nu_{\ell})$ $(\ell=e,\mu)$.
Currently, the SM prediction is roughly three standard deviations away from the global average of results 
from the $BABAR$, Belle, and LHCb experiments. 
Many theoretical efforts have been made in resolving the issue of ${\cal R}_{D^{(*)}}$ anomaly and  
searching for new physics beyond the SM~\cite{Lat1,Lat2,Bigi,LCSR1}.
In view of this,  tests of LFU in  $D$ decays are also intriguing complementary endeavors.

Exclusive semileptonic and rare $D$ decays provide rigorous tests of the SM in the charm sector
including not only the LFU but also the Cabibbo-Kobayashi-Maskawa (CKM) matrix elements~\cite{Cabibbo,Koba},
which  describe the mixings among the quark flavors in the weak decays and hold the key to the $CP$ violation issues in the quark sector.
Compared to the semileptonic  $D\to \pi(K)\ell\nu (\ell=e, \mu)$ decays 
induced by flavor-changing charged current, the rare  $D\to \pi(K)\ell\ell$  decays 
are induced by the flavor-changing neutral current (FCNC). 
Since the rare decays are loop-suppressed in the SM as they proceed through FCNC, 
they are also pertinent to test the SM and search for physics beyond the SM.
Recent BES III measurements~\cite{BES15,BES1,BES2,BES3,BES4,BES5,BES6,BES7} for many exclusive semileptonic charm decays 
also allow one to test the SM in the charm sector more precisely. 

While the experimental measurements of exclusive decays are much easier than those of inclusive ones, 
the theoretical knowledge of exclusive decays is sophisticated essentially due to the hadronic form factors entered
in the long distance nonperturbative contributions. Along with new  particle effects beyond the SM, which may amend the Wilson coefficients
of the effective weak Hamiltonian that describes physics below the electroweak scale,  the reliable and precise calculations of the hadronic form factors 
are very important to constrain the SM and search for new physics effects beyond the SM.

The calculations of hadronic form factors for semileptonic and rare $D$ decays  have been made by various theoretical approaches,
such as lattice QCD~(LQCD)~\cite{NA10,NA11,Lub17,Lub18}, QCD sum rules~\cite{BBD, CF},  QCD light-cone sum rules~\cite{Ball06,KK,Fu}, 
symmetry-preserving continuum approach to the SM strong-interaction bound-state problem~\cite{Yao20},
quark potential model~\cite{ISGW, ISGW2, IW90},
relativistic  quark model(RQM) based on the quasipotential approach~\cite{FGK20}, covariant confining quark model (CCQM)~\cite{Ivanov19},
chiral quark model~\cite{Palmer}, and constituent quark model~\cite{MS00} etc.

Perhaps, one of the most apt formulations for the analysis of exclusive processes involving hadrons may be provided in the framework of 
light-front (LF) quantization~\cite{SPP}. The semileptonic and rare $D$ decays have also been analyzed by 
the light-front quark model (LFQM)~\cite{SLF2,CK17,Cheng97,KT,QC18,QC20,CLF1,Cheng04,Verma12} based on the LF quantization.

In the standard LFQM that we use in this work, the constituent quark and antiquark
in a bound state are required to be on-mass shells and the spin-orbit wave function (WF) is
obtained by the interaction-independent Melosh transformation~\cite{Melosh} from the ordinary equal-time static spin-orbit
WF assigned by the quantum number $J^{PC}$. For the radial part, we use the phenomenologically accessible
Gaussian WF $\phi(x, {\bf k}_\perp)$.  Since the standard LFQM itself is not amenable to pin down the zero modes, 
the exactly solvable manifestly covariant Bethe-Salpeter (BS) model with the simple multipole type $q{\bar q}$ vertex  
was utilized~\cite{CLF1,Cheng04,CJ09,BCJ2} to help
identify the zero modes in a systematic way. On the other hand, this BS model is less realistic than the standard LFQM.
Thus, as an attempt to apply the zero modes found in the BS model to the standard LFQM, 
the effective replacement~\cite{CLF1,Cheng04,CJ09} of
the LF vertex function $\chi(x,{\bf k}_\perp)$ obtained in the BS model with the more 
realistic Gaussian WF $\phi(x,{\bf k}_\perp)$ in the standard LFQM has been made.

However,  we found~\cite{CJ14,CJ15,CJ17} that  the correspondence relation between $\chi$ and $\phi$
proposed in~\cite{CLF1,Cheng04,CJ09} encounters the self-consistency problem, e.g. the vector meson decay constant
obtained in the standard LFQM was found to be different
for different sets of the LF current components and polarization states of the vector meson~\cite{CJ14}.  
We also resolved~\cite{CJ14,CJ15,CJ17}  this self-consistency problem by imposing
the on-mass shell condition of the constituent quark and antiquark in addition to the original correspondence relation between $\chi$ and $\phi$.
Specifically,  our new finding for the constraint of the on-mass shell condition corresponds to the replacement of physical meson mass $M$ with the invariant mass $M_0$
in the calculation of the  matrix element.
The remarkable feature of our new additional correspondence  relation $(M\to M_0)$ between
the two models in the calculations of the two-point functions such as the weak decay constants 
and the distribution amplitudes of mesons~\cite{CJ14,CJ15,CJ17} 
was that the LF treacherous points such as the zero modes and the off-mass shell instantaneous contributions 
appeared in the BS model are absent in the standard LFQM.  This prescription $(M\to M_0)$ can be regarded as an
effective inclusion of the zero modes in the valence region of the LF calculations.

As an extension our analysis of the two-point functions~\cite{CJ14,CJ15,CJ17}  to the three-point ones, 
in our very recent LFQM analysis~\cite{Choi21} of the semileptonic $B\to D\ell\nu_\ell$ decays, we presented
the self-consistent descriptions of the weak transition form factors (TFFs) $f_+(q^2)$ and $f_-(q^2)$.  Especially,
$f_-(q^2)$ should be obtained by using least two components of the weak vector current $J^\mu_V$ while $f_+(q^2)$ can be obtained 
from the single and ``good" component ($J^+_V$) of the current.
Because of this,  $f_-(q^2)$ has been known to receive the zero mode mainly due to the unavoidable usage 
of the so called ``bad"  components of the current,  i.e. ${\bf J}_{\perp V}=(J_x, J_y)$ and $J^-_V$, 
many efforts  have been made to obtain the Lorentz covariant form factors~\cite{CLF1,Cheng04,CJ09} within the standard LFQM 
by properly handling the zero-mode as well as the instantaneous contributions. 
Applying the same correspondence relations found in~\cite{CJ14,CJ15,CJ17} to the $B\to D\ell\nu_\ell$ decays,
we found that 
the zero modes and instantaneous contributions to $f_-(q^2)$ are made to be absent in the standard LFQM while they exist in the BS model.
In other words, we obtained the current-component independent form factor $f_-(q^2)$ in the standard LFQM,
i.e.  $f_-(q^2)$ obtained from $(J^+, J^-)_V$ is exactly the same as the one obtained from 
 $(J^+, {\bf J}_\perp)_V$ numerically and both are expressed  as the convolution of the initial and final state LFWFs 
 in the valence region of the $q^+=0$ frame.
This verifies that our new correspondence relations found in the two-point functions are also applicable to the three-point functions.

The purpose of this paper is to extend our previous analysis~\cite{Choi21} of the  form
factors $f_\pm(q^2)$ for the semileptonic decays between  the two pseudoscalar mesons to obtain the 
current-component independent tensor form factor $f_T(q^2)$ for  the
rare decays, which complete the set of hadronic matrix elements regulating the exclusive semileptonic and rare
decays between the two pseudoscalar mesons.
We then apply our Lorentz covariant form factors $(f_\pm, f_T)$ 
for the analysis of  the semileptonic and rare $D\to \pi(K)$ decays 
within the standard model and 
the light-front quark model (LFQM) constrained by the variational principle for the QCD-motivated 
effective Hamiltonian~\cite{CJ09,CJ99,CJ99PLB,Choi07}.

The paper is organized as follows: In Sec.~\ref{sec:II},  we  introduce three form factors 
$(f_\pm, f_T)$ for the
semileptonic and rare decays between two pseudoscalar mesons. In the $q^+=0$ frame, we define
the form factors extracted from the various combinations of vector and tensor currents.
In Sec.~\ref{sec:III}, we set up the current matrix elements for the form factors in
an exactly solvable model based on the covariant BS model of ($3+1$) dimensional fermion field theory.
We then present our LF calculations of tensor form factor $f_T$ in the BS model using the two different sets 
%all three components $(J^\pm, {\bf J}_\perp)_V$ of the weak vector current $J^\mu_V$ and 
($J^{+\perp}_T$ and $J^{+-}_T$) of the tensor current $J^{\mu\nu}_T$. For completeness, we also
present the results of the current-component independent form factors $f_\pm(q^2)$ found ~\cite{Choi21}.
We note that while $f_T(q^2)$ obtained from $J^{+\perp}_T$ is immune to the zero mode and the instantaneous contribution,
$f_T(q^2)$ obtained from $J^{+-}_T$ cannot avoid those contributions in this BS model.
Linking the covariant BS model to the standard LFQM with our new correspondence relations
between the two models~\cite{CJ14,CJ15,CJ17},  however, we 
find that $f_T(q^2)$ obtained from $J^{+-}_T$ in the standard LFQM is made to be free of the zero mode as well as the instantaneous contribution.
In other words, we obtained the current-component independent tensor form factor $f_T(q^2)$ in the standard LFQM
regardless of using $J^{+\perp}_T$ or $J^{+-}_T$ as in the case of $f_-(q^2)$ calculation~\cite{Choi21}.
Finally, we present the current-component independent TFFs $(f_\pm, f_T)$ in the $q^+=0$ frame of the standard LFQM. 
In Sec.~\ref{sec:IV}, we present our numerical results of the form factors for the semileptonic and rare $D\to \pi(K)$  decays
as well as the branching ratios for the semileptonic $D\to \pi(K)\ell\nu_\ell$ $(\ell=e,\mu)$.
Summary and discussion follow in Sec.~\ref{sec:V}.

%\section{ Semileptonic Decays between two pseudoscalar mesons}
\section{Theoretical Framework}
\label{sec:II}
The matrix elements of the vector $J_V^\mu={\bar q}\gamma^\mu c$ and the tensor 
$J_T^{\mu\nu}={\bar q}\sigma^{\mu\nu} c$ currents for the weak  $c\to q(q=u,d,s)$ transitions
between the initial $D$ meson and the final $\pi$ or $K$ meson can be parametrized by the following set of 
invariant form factors, ($f_+, f_-, s$)~\cite{IW90}:
\be\label{eq1c}
{\cal M}^\mu_{V}\equiv\la P_2|J_V^\mu|P_{1}\ra =  f_{+}(q^2) P^{\mu} +
f_-(q^2)q^\mu,
 \ee
 and 
 \be\label{eq2c}
{\cal M}^{\mu\nu}_{T}\equiv\la P_2|J_T^{\mu\nu}|P_{1}\ra =  is (q^2) [ P^\mu q^\nu - q^\mu P^\nu],
 \ee
where $P=P_1+P_2$ and $q = P_1-P_2$ is the
four-momentum transfer to the lepton pair($\ell\nu_\ell$) with $m^2_\ell\leq q^2\leq (M_1-M_2)^2$ for the
semileptonic decays or to the pair ($\ell^+\ell^-$) with $4m^2_\ell\leq q^2\leq (M_1-M_2)^2$ for the rare decays,
respectively.  The antisymmetric tensor in Eq.~(\ref{eq2c}) is given by
 $\sigma^{\mu\nu} = (i/2)[\gamma^\mu, \gamma^\nu]$.
 
On many occasions, it is useful to express Eq.~(\ref{eq1c}) in terms of the form factors $f_{+}(q^2)$ and $f_0(q^2)$, which are
related to the transition amplitude with the exchange of a vector ($1^-$) and a scalar ($0^+$) boson in the $t$-channel, respectively, and satisfy
\be\label{eq3c}
f_0(q^2) = f_+(q^2) +
\frac{q^2}{M^2_1-M^2_2}f_-(q^2).
\ee
Likewise, the tensor form factor $s(q^2)$ in Eq.~(\ref{eq2c}) can also be redefined by
\be\label{eq4c}
s(q^2) = \frac{f_T(q^2)}{M_1 + M_2},
\ee
to make $f_T(q^2)$ dimensionless.

Including the nonzero lepton mass ($m_\ell$), the differential decay rate for the
semileptonic $P_1\to P_2\ell\nu_\ell$ process is given by~\cite{KS, Yao}
\bea\label{eq5c} 
\frac{d\Gamma}{dq^{2}}&=&
 \frac{G^{2}_{F}}{96\pi^{3}} |{\vec p}^*| |V_{Q_{1}\bar{Q}_{2}}|^{2}\frac{q^2}{M^2_1}
\biggl(1-\frac{m^2_\ell}{q^2}\biggr)^2
\nonumber\\
&&\times
\left[
\biggl(1+\frac{m^2_\ell}{2q^2}\biggr) |H_+|^{2}
+
\frac{3 m^2_\ell}{2 q^2}|H_0|^2
\right],
\eea
where $G_{F}=1.166\times 10^{-5}$ GeV$^{-2}$ is the Fermi constant, $V_{Q_{1}\bar{Q}_{2}}$ is the relevant CKM
mixing matrix element, and
\be\label{eq6c} 
|{\vec p}^*| = \frac{1}{2M_{1}}\sqrt{
(M_{1}^{2}+M_{2}^{2}-q^{2})^{2}-4M_{1}^{2}M_{2}^{2}}
 \ee
 is the modulus of the three-momentum of the daughter meson in the parent meson rest frame and the helicity amplitudes
 $H_+$ and $H_0$ are given by
 \be\label{eq7c}
 H_+ = \frac{2 M_1 |{\vec p}^*| }{\sqrt{q^2}} f_{+}(q^2),\;\;
  H_0 = \frac{M^2_1- M^2_2 }{\sqrt{q^2}} f_{0}(q^2).
 \ee
We note that $q^2=q^2_{\rm max}$
corresponds to zero-recoil of the final meson in the initial meson rest frame and
the $q^2 =0$ corresponds to the maximum recoil of the final
meson recoiling with the maximum three momentum $|{\vec P}_2|=\frac{(M^2_1 - M^2_2)}{2 M_1}$. 

In the LF calculation of the form factors, we use the metric  convention $a\cdot b =\frac{1}{2} (a^+ b^- + a^- b^+) - {\bf a}_\perp\cdot {\bf b}_\perp$.
Performing the LF calculation in the $q^+=0$ frame  (i.e. $q^2=-{\bf q}^2_\perp=-Q^2<0$) with $P_1=(P^+_1, P^-_1, {\bf P}_{1\perp})=(P^+_1, M^2_1/P^+_1,{\bf 0}_\perp)$,
we utilize all three components ($\mu, \nu=+, -, \perp$) of the current $J^\mu_V$ and $J^{\mu\nu}_T$ in Eqs.~(\ref{eq1c}) and (\ref{eq2c})
to obtain  $f_+(q^2)$,  $f_-(q^2)$ [or $f_0(q^2)$], and $f_T(q^2)$.
The form factors obtained in the spacelike region $(q^2 < 0)$ are then analytically continued to the timelike region by changing ${\bf q}^2_\perp$
to $-q^2$ in the form factors as we show in our numerical calculations.

While the form factor $f_+(q^2)$ can be obtained from the plus component ($J^+_V$) of the vector current, one cannot but use
two different combinations of the current to obtain $f_-(q^2)$ such as 
$(J^+, {\bf J}_{\perp})_V$  or $(J^+, J^-)_V$. 
That is, using those sets of the current components in the $q^+=0$ frame, one obtains the relations between 
the weak form factors $f_\pm(q^2)$ and the current
matrix elements in Eq.~(\ref{eq1c}) as follows~\cite{Choi21}:
 \bea\label{eq8c}
 f_+(q^2) &=& \frac{{\cal M}^+_V}{2P^+_1},
%\nonumber
\\
 f^{(+\perp)}_-(q^2) &=&  \frac{{\cal M}^+_V}{2P^+_1}+ \frac{ {\cal M}^\perp_V \cdot {\bf q}_\perp}{ {\bf q}^2_\perp},
% \nonumber
\\
 f^{(+-)}_-(q^2) &=& -  \frac{{\cal M}^+_V}{2P^+_1} \biggl( \frac{\Delta M^2_{+}  + {\bf q}^2_\perp}{\Delta M^2_{-}  - {\bf q}^2_\perp} \biggr)
  + \frac{ P^+_1 {\cal M}^-_V}{ \Delta M^2_{-}  - {\bf q}^2_\perp},
 \eea
where $\Delta M^2_{\pm} = M^2_1 \pm M^2_2$ and we denote $f_-(q^2)$ obtained from $(J^+, {\bf J}_\perp)_V$ and $(J^+, J^-)_V$
as $f^{(+\perp)}_-(q^2)$ and $ f^{(+-)}_-(q^2)$, respectively. 
It is prerequisite to show that  $f^{(+\perp)}_-(q^2)=f^{(+-)}_-(q^2)$ to assert the Lorentz invariance of the form factor and the self-consistency of the
model.

Likewise, the tensor form factor $s(q^2)$ can be obtained from using either 
$J^{+\perp}_T$ or $J^{+-}_T$. In this case, the relations between $s(q^2)$ and the current matrix element in Eq.~(\ref{eq2c}) 
are given by
 \bea\label{eq9c}
 s^{(+\perp)} (q^2) &=& - \frac{i {\cal M}^{+\perp}_T \cdot {\bf q}_\perp}{2 {\bf q}^2_\perp P_1^+},\\
% \nonumber\\
  s^{(+-)} (q^2)&=& - \frac{i {\cal M}^{+-}_T }{2 (\Delta M^2_{-}  - {\bf q}^2_\perp)},
 \eea
where $s^{(+\perp)}(q^2)$ and $ s^{(+-)}(q^2)$ represent the form factor $s(q^2)$ obtained 
from $J^{+\perp}_T$ and $J^{+-}_T$, respectively.  Of course, $s^{(+\perp)}(q^2)= s^{(+-)}(q^2)$ should be satisfied in the
self-consistent model calculation.

Our aim in this work is to show 
$s^{(+\perp)}(q^2)= s^{(+-)}(q^2)$ in addition to our previous verification of  $f^{(+\perp)}_-(q^2)=f^{(+-)}_-(q^2)$~\cite{Choi21} in our LFQM,
which completes the analysis of the exclusive semileptonic and rare decays between two pseudoscalar mesons.
For this purpose, we start from the exactly solvable manifestly covariant BS model and then connect it to our phenomenologically
accessible LFQM.
Although we analyzed $f_\pm(q^2)$ in~\cite{Choi21}, we shall include them again in the next section for the completeness of the analysis.

\section{Model Description}
\label{sec:III}
\subsection{Manifestly covariant model}
In the solvable model, based on the covariant BS model of
($3+1$)-dimensional fermion field theory~\cite{BCJ2,Choi21}, the matrix elements
${\cal M}=({\cal M}^\mu_V$, ${\cal M}^{\mu\nu}_T)$
%${\cal M}^\mu_{V}$ and ${\cal M}^{\mu\nu}_{T}$
 are given by
 \be\label{eq10c}
{\cal M} = iN_c\int\frac{d^4k}{(2\pi)^4} \frac{H_{p_1}{\cal T} H_{p_2}} {N_{p_1} N_{k} N_{p_2}},
 \ee
% and
% \be\label{eq11c}
%  {\cal M}^{\mu\nu}_{T} = iN_c\int\frac{d^4k}{(2\pi)^4} \frac{H_{p_1}T^{\mu\nu} H_{p_2}} {N_{p_1} N_{k} N_{p_2}},
% \ee
where 
%${\cal A}=({\cal M}^\mu_V$, ${\cal M}^{\mu\nu}_T)$ pairs with 
the corresponding trace terms ${\cal T}=(S^\mu, T^{\mu\nu})$ are given by
%the trace terms are given by
\be\label{eq11c}
S^\mu  = {\rm Tr}[\gamma_5\left(\slash \!\!\!\!\! p_1 +m_1 \right) \gamma^\mu \left(\slash \!\!\!\!\! p_2 +m_2 \right)\gamma_5
 \left(-\slash \!\!\!\! k + m_q \right)],
 \ee
for the vector current and
 \be\label{eq12c}
T^{\mu\nu}  = {\rm Tr}[\gamma_5\left(\slash \!\!\!\!\! p_1 +m_1 \right)  \sigma^{\mu\nu} \left(\slash \!\!\!\!\! p_2 +m_2 \right)\gamma_5
 \left(-\slash \!\!\!\! k + m_q \right)],
 \ee
for the tensor current, respectively. $N_c$ is the number of colors and 
$p_j =P_j -k (j=1,2)$ and $k$ are the internal momenta carried by the quark and antiquark propagators
of mass $m_j$ and $m_q$, respectively. The corresponding denominators are given by  $N_{p_j} = p^2_{j} - m^2_j + i\ep$ and
$N_{k} = k^2 - m^2_q + i\ep$. We take
the $q\bar{q}$ bound-state vertex functions $H_{p_j}(p^2_j,k^2)=g_j/(p^2_j -\Lambda^2_j+i\epsilon)$
of the initial ($j=1$) and final ($j=2$) state pseudoscalar mesons, where $g_j$ and $\Lambda_j$ are
constant parameters in this manifestly covariant model.

Performing the LF calculation in the $q^+=0$ frame, one obtains the  following identity $\not\!\!q = \not\!\!q_{\rm on}  +\frac{1}{2}  \gamma^+\Delta^-_q$,
where $\Delta^-_q = q^- - q^-_{\rm on}$ and 
the subscript (on) denotes the on-mass shell  quark momentum,
i.e., $p^2_{j\rm on}=m^2_j$ and $k^2_{\rm on}=m^2_q$.
Using this identity,  one can separate the trace terms into the ``on"-mass shell propagating part
and the ``off"-mass shell instantaneous one, i.e.
$S^\mu= S^\mu_{\rm on} + S^\mu_{\rm off}$ for the vector current and
$T^{\mu\nu}= T^{\mu\nu}_{\rm on} + T^{\mu\nu}_{\rm off}$ for the tensor current.
%Within this manifestly covariant BS model, one can explicitly find that the perpendicular and minus components of the LF current
%receive in general  the zero modes and the off-mass shell instantaneous contributions, while the plus component
%of the current does not receive them.

The explicit LF calculation in parallel with the manifestly covariant calculation of Eq.~(\ref{eq10c}) to compute
 $f_\pm (q^2)$ can be found in~\cite{CJ09} where $f_-(q^2)$ was obtained from $f^{(+\perp)}_-(q^2)$. The identical 
 result for $f^{(+\perp)}_-(q^2)$ was also obtained in~\cite{CLF1,Cheng04} using the so called ``covariant LFQM" analysis.
As shown in Ref.~\cite{CLF1,Cheng04,CJ09}, while $f_{+}(q^2)$ obtained from the plus current was immune to the zero mode, 
the form factor $f_-(q^2)$ received both instantaneous and zero-mode contributions. The same situation happens for
$f^{(+-)}_-(q^2)$ although the zero-mode and the instantaneous contributions may differ quantitatively from $f^{(+\perp)}_-(q^2)$.
However, as we have shown in~\cite{Choi21}, $f^{(+\perp)}_-(q^2)$ and $f^{(+-)}_-(q^2)$ obtained in the standard LFQM by using our new correspondence
relations between the BS model and the standard LFQM show identical result in the valence region of the $q^+=0$ frame
without involving explicit zero modes and instantaneous contributions. 
In this work, we shall show that the tensor form factor $s(q^2)$ in the standard LFQM is 
independent of the components of the current, i.e. $s^{(+\perp)}(q^2)=s^{(+-)}(q^2)$. We should note that all of those equalities,
i.e.  $f^{(+\perp)}_-(q^2)=f^{(+-)}_-(q^2)$ and $s^{(+\perp)}(q^2)=s^{(+-)}(q^2)$ are derived from
the constraint of the on-mass shellness of the quark and antiquark propagators together with the zero-binding energy limit (i.e. $M=M_0$)
used in the standard LFQM.

Therefore, from now on, we discuss only for the on-mass shell contributions in the valence region of the $q^+=0$ frame 
between the manifestly covariant BS model and the standard LFQM.
The on-shell contributions to $S^\mu$ and $T^{\mu\nu}$ are given by
\bea\label{eq13c}
S^\mu_{\rm on} &=&
4 \biggl[
 p^\mu_{1\rm on} (p_{2\rm on}\cdot k_{\rm on} + m_2m_{\bar q} ) + p^\mu_{2\rm on} (p_{1\rm on}\cdot k_{\rm on} +  m_1m_{\bar q})
 \nonumber\\
&&~~~ +    k^\mu_{\rm on} (m_1 m_2 - p_{1\rm on}\cdot p_{2\rm on})
\biggr],
 \eea
 and
 \bea\label{eq14c}
T^{\mu\nu}_{\rm on} &=&
4 i \biggl[
p_{1\rm on}^{\mu }  (m_2   k_{\rm on}^{\nu }+ m_{\bar q}  p_{2\rm on}^{\nu } )
- p_{2\rm on}^{\mu } ( m_1 k_{\rm on}^{\nu } + m_{\bar q}  p_{1\rm on}^{\nu } )
\nonumber\\
&&+ k_{\rm on}^{\mu } (m_1   p_{2\rm on}^{\nu } -  m_2 p_{1\rm on}^{\nu } )
\biggr],
 \eea
respectively. The LF four-momenta of the on-mass shell quark and antiquark propagators 
in the $q^+=0$ (i.e. $P^+_1=P^+_2$) frame are given by
\bea\label{eq15c}
p_{1\rm on}&=& \left[ x P^+_1, \frac{m^2_1 + {\bf k}_\perp^2}{xP^+_1}, -{\bf k}_\perp \right],\\
%\nonumber\\
p_{2\rm on} &=& \left[ x P^+_1, \frac{m^2_2 + ({\bf k}_\perp+{\bf q}_\perp)^2}{xP^+_1}, -{\bf k}_\perp-{\bf q}_\perp \right],\\
%\nonumber\\
k_{\rm on} &=& \left[ (1-x) P^+_1, \frac{m^2_q + {\bf k}_\perp^2}{ (1-x)P^+_1}, {\bf k}_\perp \right],
\eea
where 
$x=\frac{p^+_1}{P^+_1}$ and ${\bar x}= \frac{k^+}{P^+_1}$ are the LF longitudinal momentum fractions of the quark and antiquark, which
satisfy $x+ {\bar x}=1$.

By the integration over $k^-$ in Eq.~(\ref{eq10c}) and closing the contour in the lower half of the complex $k^-$ plane, one
picks up the residue at $k^-=k^-_{\rm on}$ in the valence region of $0<k^+<P^+_2$ (or $0<x<1$). 
We denote the on-mass shell contribution to ${\cal M}^\mu_V$ and ${\cal M}^{\mu\nu}_T$ in the valence region
as $[{\cal M}^\mu_V]^{\rm BS}_{\rm on}$ and $[{\cal M}^{\mu\nu}_T]^{\rm BS}_{\rm on}$, respectively.
The explicit forms of $[{\cal M}^\mu_V]^{\rm BS}_{\rm on}$ and $[{\cal M}^{\mu\nu}_T]^{\rm BS}_{\rm on}$
are obtained as~\cite{Choi21}
%Then the Cauchy integration formula for the
%$k^-$ integration in the valence region of Eqs.~(\ref{eq10c}) and (\ref{eq11c}) yields
  \be\label{eq16c}
 {\cal M}^{\rm BS}_{\rm on} =
N_c\int^1_0 \frac{dx}{\bar x}\int \frac{d^2{\bf k}_\perp}{16\pi^3}
\chi_1(x,{\bf k}_\perp) \chi_2 (x, {\bf k'}_\perp)
{\cal T}_{\rm on} ,
  \ee
where ${\cal M}=({\cal M}^\mu_V$, ${\cal M}^{\mu\nu}_T)$ pairs with ${\cal T}=(S^\mu, T^{\mu\nu})$.
The LF quark-meson vertex function $\chi_{1(2)}$ of the initial (final) state is given by 
\be\label{eq17c}
\chi_{1(2)}  (x, {\bf k^{(\prime)}}_\perp)= \frac{g_{1(2)}}{x^2 (M^2_{1(2)} - M^{(\prime)2}_0)(M^2_{1(2)} -M^{(\prime)2}_{\Lambda_1{(\Lambda_2)}})},
\ee
where ${\bf k'}_\perp={\bf k}_\perp + (1-x) {\bf q}_\perp$ and
\bea\label{eq18c}
M^2_0 &=& \frac{{\bf k}^2_\perp + m^2_1}{x} + \frac{{\bf k}^2_\perp + m^2_q}{ 1-x},\\
%\nonumber\\
M'^2_0 &=& \frac{{\bf k'}^2_\perp + m^2_2}{x} + \frac{{\bf k'}^2_\perp + m^2_q}{ 1-x},
\eea
are the invariant masses of the initial and final states, respectively. Likewise, $M_{\Lambda_{1(2)}}$ are obtained as
$M_{\Lambda_1} = M_0 (m_1\to\Lambda_1)$ and $M'_{\Lambda_2} = M'_0 (m_2\to\Lambda_2)$.
%$M^2_{0(\Lambda_1)}= \frac{{\bf k}^2_\perp + m^2_1(\Lambda^2_1)}{x} + \frac{{\bf k}^2_\perp + m^2_q}{ 1-x}$
%and $M'^2_{0(\Lambda_2)}=M^2_{0(\Lambda_1)}(m_1(\Lambda_1)\to m_2(\Lambda_2),  {\bf k}_\perp\to {\bf k'}_\perp={\bf k}_\perp + (1-x) {\bf q}_\perp$).

For the trace ${\cal T}=(S^\mu, T^{\mu\nu})$ calculations relevant to the form factors,  
the on-mass shell contributions $S^\mu_{\rm on}$ obtained from all three components $\mu=(+, \perp, -)$ of the vector current $J^\mu_V$
are given by~\cite{Choi21}
 \bea\label{eq19c}
 S^+_{\rm on} &=& \frac{4 P^+_1}{{\bar x}} ({\bf k}_\perp\cdot{\bf k'}_\perp
  + {\cal A}_1{\cal A}_2 ),
 \eea
%\nonumber\\
\bea
{\bf S}_{\perp\rm on}
&=& \frac{-2 {\bf k}_\perp}{x {\bar x}} 
\biggl [
 2{\bf k}_\perp\cdot{\bf k'}_\perp + {\bar x} ({\bf q}^2_\perp + m^2_1 + m^2_2)  + 2x^2 m^2_{q}
\nonumber\\
 &&
 + 2x {\bar x} (m_1m_{q} + m_2m_{q} - m_1m_2) 
 \biggr]
 -\frac{2{\bf q}_\perp}{x {\bar x}} ( {\bf k}^2_\perp + {\cal A}^2_1),
\nonumber\\
\eea
\bea
S^-_{\rm on}
&=& \frac{4}{x^2 {\bar x} P^+_1} 
\biggl [
{\bar x} (m_1 {\cal A}_1 + {\bf k}^2_\perp) [m^2_2 + ({\bf k}_\perp + {\bf q}_\perp)^2]
\nonumber\\
&&
+ x^2 {\bar x} M^2_0 ({\bf k}^2_\perp + {\bf k}_\perp\cdot{\bf q}_\perp)
+ x^2 m_1 m_2 (m^2_{q} + {\bf k}^2_\perp) 
\nonumber\\
&&+ x {\bar x} m_2 m_{q} (m^2_1 + {\bf k}^2_\perp)
\biggr],
 \eea
 where ${\cal A}_j= (1-x) m_j + x m_{q} (j=1,2)$.
 Likewise, the on-shell contributions $T^{\mu\nu}_{\rm on}$  obtained from the two sets of the tensor current $J^{\mu\nu}_T$, i.e.
 $(\mu,\nu) =(+, \perp)$ and $(+, -)$,  are given by

 \bea\label{eq20c}
 T^{+\perp}_{\rm on} &=& -4 i P^+_1  \left[ (m_1 - m_2) {\bf k}_\perp + {\cal A}_1 {\bf q}_\perp \right],\\
% \nonumber\\
% {\cal T}^{+-}_{\rm on\sigma} &=& -2 (M^2_1 - M^2_2) \left[ (m_1 - m_2) {\bf k}_\perp + {\cal A}_1 {\bf q}_\perp \right],
%\eea
%\begin{widetext}
% \bea
T^{+-}_{\rm on} &=& \frac{4 i}{x{\bar x}}
\biggl[
(1- 2x) (m_1 - m_2) {\bf k}^2_\perp + 2 {\bar x} {\cal A}_1 {\bf k}_\perp\cdot{\bf q}_\perp
\nonumber\\
&&+ {\bar x} {\cal A}_1 {\bf q}^2_\perp + (m_2 - m_1) {\cal A}_1 {\cal A}_2
\biggr].
 \eea
% \end{widetext}
Using Eqs.~(8)-(12) and~(\ref{eq16c}), one obtains the on-mass shell contributions 
to the weak form factors $(f_+, f_-, s)$ as follows 
\be\label{eq21c}
{\cal F}^{\rm BS}_{\rm on} = N_c \int^{1}_{0}\frac{dx}{\bar x} \int \frac{d^{2}{\bf k}_{\perp}}{16\pi^3}
\chi_{1}(x,{\bf k}_{\perp})\la {\cal O}\ra_{\rm on}^{\rm BS}  \chi_{2}(x,{\bf k'}_{\perp}),
\ee
where the form factors ${\cal F}=\{ f_+,  f^{(+\perp)}_-, f^{(+-)}_-, s^{(+\perp)}, s^{(+-)} \}$ obtained from different combinations of the vector and tensor
currents pair with the following corresponding operators
$\la {\cal O}\ra_{\rm on}^{\rm BS}=\{ {\cal O}_+,  {\cal O}^{(+\perp)}_-, {\cal O}^{(+-)}_-, {\cal O}^{(+\perp)}_s, {\cal O}^{(+-)}_s \}$
including the spin and external momenta factors:
\bea\label{eq22c}
{\cal O}_+ &=& \frac{ S^{+}_{\rm on}} {2P^+_1},\\
%\nonumber\\
 {\cal O}^{(+\perp)}_- &=& \frac{S^{+}_{\rm on}} {2P^+_1} + \frac{{\bf S}_{\perp\rm on}\cdot{\bf q}_{\perp}}{{\bf q}^2_\perp},\\
% \nonumber\\
{\cal O}^{(+-)}_- &=& -  \frac{S^+_{\rm on}}{2P^+_1} \biggl( \frac{\Delta M^2_{+}  + {\bf q}^2_\perp}{\Delta M^2_{-}  - {\bf q}^2_\perp} \biggr)
  + \frac{ P^+_1 S^-_{\rm on}}{ \Delta M^2_{-}  - {\bf q}^2_\perp},\\
%\nonumber\\
 {\cal O}^{(+\perp)}_s &=& - \frac{i T^{+\perp}_{\rm on} \cdot {\bf q}_\perp}{2 {\bf q}^2_\perp P_1^+},\\
% \nonumber\\
  {\cal O}^{(+-)}_s &=& - \frac{i T^{+-}_{\rm on} }{2 (\Delta M^2_{-}  - {\bf q}^2_\perp)}.
\eea
%%%%%%%%%

In the manifestly covariant BS model given by Eq.~(\ref{eq10c}), we note that only the two form factors
$f_+(q^2)$ and $s^{(+\perp)}(q^2)$ defined by Eqs.~(\ref{eq22c}) and (34), respectively, are 
are immune to the zero modes as well as the instantaneous contributions and thus exactly
equal to the full exact solution (i..e. manifestly covariant solution), i.e.
$[f_+]^{\rm BS}_{\rm on}=f^{\rm Cov}_+$  and $[s^{(+\perp)}]^{\rm BS}_{\rm on}=s^{\rm Cov}$, 
without involving such LF treacherous points.
However, since the other three form factors 
$f^{(+\perp)}_-$, $f^{(+-)}_-$, and $s^{(+-)}$
are contaminated by the zero modes as well as the  instantaneous contributions, the on-mass shell contributions
$[f^{(+\perp)}_-]^{\rm BS}_{\rm on}$, $[f^{(+-)}_-]^{\rm BS}_{\rm on}$, and $[s^{(+\perp)}]^{\rm BS}_{\rm on}$ themselves
can never be the same as the exact solutions unless the zero modes and the instantaneous contributions
are taken into account.
Furthermore, one can easily check that $[f^{(+\perp)}_-]^{\rm BS}_{\rm on} \neq [f^{(+-)}_-]^{\rm BS}_{\rm on}$ and
$[s^{(+-)}]^{\rm BS}_{\rm on}\neq [s^{(+\perp)}]^{\rm BS}_{\rm on}$.

However, in the following subsection, we shall show in the standard LFQM (denoted by SLF) that 
$f^{\rm SLF}_-=[f^{(+\perp)}_-]^{\rm SLF}_{\rm on} = [f^{(+-)}_-]^{\rm SLF}_{\rm on}$ and
$s^{\rm SLF}=[s^{(+-)}]^{\rm SLF}_{\rm on} = [s^{(+\perp)}]^{\rm SLF}_{\rm on}$ without involving
explicit zero-mode and instantaneous contributions, which comes after using our new correspondence 
relations between the BS model and the standard LFQM.

\subsection{The standard LFQM}
\label{sec:IIIb}
In the standard LFQM~\cite{SLF2,Cheng97,KT,CJ98,CJ99,CJ99PLB,Choi07,CJ07},
%Cheng97,CCP,Card95,MFPS02,CJK02,CJ_kaon,CJ06,CJ08,CJ_Bc,CJ_GPD01,CYA}, 
the LF wave function (LFWF) of a ground state pseudoscalar meson
as a $q\bar{q}$ bound state is given by
\be\label{eq23c}
\Psi_{\lam{\bar\lam}}(x,{\bf k}_{\perp})
={\phi(x,{\bf k}_{\perp})\cal R}_{\lam{\bar\lam}}(x,{\bf k}_{\perp}),
\ee
where ${\cal R}_{\lam{\bar\lam}}(x,{\bf k}_{\perp})$ is the spin-orbit WF
that is obtained by the interaction-independent Melosh transformation from the ordinary
spin-orbit WF assigned by the quantum number $J^{PC}$.
The covariant form of ${\cal R}_{\lam{\bar\lam}}$ with the definite spin $(S, S_z)=(0,0)$
constructed out of the LF helicity $\lam({\bar\lam})$ of a quark (antiquark)
is given by
\be\label{eq24c}
{\cal R}_{\lam{\bar\lam}}
=\frac{\bar{u}_{\lam}(p_q)\gamma_5 v_{{\bar\lam}}( p_{\bar q})}
{\sqrt{2}[M^{2}_{0}-(m_1 -m_{q})^{2}]^{1/2}},
\ee
which satisfies the unitarity condition,
$\sum_{\lam{\bar\lam}}{\cal R}_{\lam{\bar\lam}}^{\dagger}{\cal R}_{\lam{\bar\lam}}=1$.
Its explicit  matrix form is given by
%}
\be\label{eq25c}
{\cal R}_{\lam{\bar\lam}}
=\frac{1}{\sqrt{2}\sqrt{{\bf k}^2_\perp+{\cal A}^2_1}}
\begin{pmatrix}
-k^L &  {\cal A}_1\\ -{\cal A}_1 & -k^R
\end{pmatrix},
\ee
where $k^{R} = k_x + i k_y$ and $k^{L} = k_x - i k_y$.

For the radial WF $\phi(x,{\bf k}_{\perp})$ in~Eq.~(\ref{eq23c}), we use the Gaussian WF
\be\label{eq26c}
\phi(x,{\bf k}_{\perp})=
\frac{4\pi^{3/4}}{\beta^{3/2}} \sqrt{\frac{\partial
k_z}{\partial x}} {\rm exp}(-{\vec k}^2/2\beta^2),
\ee
where $\vec{k}^2={\bf k}^2_\perp + k^2_z$ and $\beta$ is the variational parameter.
%fixed by the analysis of the ground state pseudoscalar and vector meson mass spectra~\cite{CJ09,CJ99,CJ99PLB,Choi07}.
The longitudinal component $k_z$ is defined by $k_z=(x-\frac{1}{2})M_0 +
\frac{(m^2_{q}-m^2_1)}{2M_0}$, and the Jacobian of the variable transformation
$\{x,{\bf k}_\perp\}\to {\vec k}=({\bf k}_\perp, k_z)$ is given by
%\be\label{QM3}
$\frac{\partial k_z}{\partial x}
= \frac{M_0}{4 x (1-x)}[ 1-
(\frac{m^2_1 - m^2_q}{M^2_0})^2]$.
%\ee
%
The normalization of our Gaussian radial WF is then given by
\be\label{eq27c}
\int^1_0 dx \int \frac{d^2{\bf k}_\perp}{16\pi^3}
|\phi(x,{\bf k}_{\perp})|^2=1.
\ee
In particular, the key idea in our LFQM~\cite{CJ09,CJ99,CJ99PLB,Choi07}
 for mesons is to treat $\phi(x,{\bf k}_{\perp})$ as a trial function for the variational principle to the QCD-motivated 
effective Hamiltonian saturating the Fock state expansion by the constituent quark and antiquark.
Using this Hamiltonian, we analyze the meson mass spectra and various wave-function-related observables, such as decay constants, 
electromagnetic form factors of mesons in a spacelike region, and the weak form factors for the exclusive semileptonic  and rare decays of pseudoscalar 
mesons in the timelike region~\cite{CJ09,CJ99,CJ99PLB,Choi07,CJ14,CJ15,CJ17,Choi21}.
%We should note that the essential property of the standard LFQM is the on-mass shellness of the quark and antiquark propagators.

In this standard LFQM, the matrix elements of the vector and tensor currents  in Eqs.~(\ref{eq1c}) and (\ref{eq2c}) 
are obtained by the convolution formula of the initial and final state LFWFs in the $q^+=0$ frame
as follows:
 \bea\label{eq28c}
 {\cal M}^{\rm SLF}_{\rm on} &=& \sum_{\lambda's}
\int^1_0 dx \int \frac{d^2{\bf k}_\perp}{16\pi^3} \phi_1(x,{\bf k}_\perp) \phi_2 (x, {\bf k'}_\perp)
\nonumber\\
&&\times
{\cal R}^\dagger_{\lam_2{\bar\lam}}
\frac{{\bar u}_{\lam_2}(p_2)}{\sqrt{p^+_2}}\Gamma
\frac{u_{\lam_1}(p_1)}{\sqrt{p^+_1}}
{\cal R}_{\lam_1{\bar\lam}},
  \eea
where ${\cal M}=({\cal M}^\mu_V$, ${\cal M}^{\mu\nu}_T)$ pairs with $\Gamma=(\gamma^\mu, \sigma^{\mu\nu})$.

Then, we first compute the  zero-mode free form factors, i.e.
$[f_+]^{\rm SLF}_{\rm on}$ and $[s^{(+\perp)}]^{\rm SLF}_{\rm on}$, 
in the SLF formulation as follows
\be\label{eq29c}
{\cal F}^{\rm SLF}_{\rm on} = \int^{1}_{0} {\bar x} dx \int \frac{d^{2}{\bf k}_{\perp}}{32\pi^3}
\frac{\phi_{1}(x,{\bf k}_{\perp})}{\sqrt{ {\cal A}_{1}^{2} + {\bf k}^{2}_{\perp}}}
\la {\cal O}\ra_{\rm on}^{\rm SLF}
\frac{\phi_{2}(x,{\bf k}'_{\perp})}{\sqrt{ {\cal A}_{2}^{2}+ {\bf k}^{\prime 2}_{\perp}}},
\ee
where the form factors ${\cal F}=\{ f_+,  s^{(+\perp)} \}$ pair with the following corresponding operators
$\la {\cal O}\ra_{\rm on}^{\rm SLF} =\la {\cal O}\ra_{\rm on}^{\rm BS} =\{ {\cal O}_+,  {\cal O}^{(+\perp)}_s \}$ 
given by Eqs.~(\ref{eq22c}) and (34).

We should note that the main differences between the covariant BS model and the standard LFQM are attributed to the different 
spin structures of the $q{\bar q}$ system (i.e. off-shellness in the BS model vs on-shellness in the standard LFQM) and the different meson-quark vertex 
functions ($\chi$ vs. $\phi$). In other words, while the results of the covariant BS model allow the nonzero binding 
energy $E_{\rm B.E} = M^2 - M^2_0$, the SLF result is obtained from the condition of on-mass shell quark 
and antiquark (i.e. $M\to M_0$).

Comparing these two form factors ${\cal F}=\{ f_+,  s^{(+\perp)} \}$ defined in 
Eq.~(\ref{eq21c}) in the BS model and Eq.~(\ref{eq29c}) in the standard LFQM, 
one can easily find the correspondence relation between the two models as follows:
\be\label{eq30c}
\sqrt{2N_c} \frac{ \chi_{1(2)}(x,{\bf k}^{(\prime)}_\perp) } {1-x}
 \to \frac{ \phi_{1(2)}(x,{\bf k}^{(\prime)}_\perp) } {\sqrt{ {\cal A}^2_{1(2)} + {\bf k}^{(\prime)2}_\perp }}.
 \ee
%We denote Eq.~(\ref{eq30c})  as “Type I” correspondence between the two models.

In many previous LFQM analyses~\cite{CK17,CLF1,Cheng04,CJ09}, the correspondence in Eq.~(\ref{eq30c}) 
has also been used for the mapping of other physical observables 
contaminated by the zero modes and/or the instantaneous contributions.
However, applying Eq.~(\ref{eq30c}) together with the same operators given by Eqs.~(32), (33), and (35)
to the other form factors ${\cal F}=\{f^{(+\perp)}_-, f^{(+-)}_-, s^{(+-)} \}$ obtained from the
only on-mass shell contributions, one encounters the same problems as the BS model, i.e.
 $[f^{(+\perp)}_-]^{\rm SLF}_{\rm on} \neq [f^{(+-)}_-]^{\rm SLF}_{\rm on}$ and
$[s^{(+-)}]^{\rm SLF}_{\rm on}\neq [s^{(+\perp)}]^{\rm SLF}_{\rm on}$
implying that
the same physical quantities obtained from different components of the current yield different results.
%This self-inconsistency problem of the model calculations is understood since ${\cal F}=\{f^{(+\perp)}_-, f^{(+-)}_-, s^{(+-)} \}$
%are known to receive the zero modes as well as the instantaneous contributions as shown in the manifestly
%covariant BS model.

In our previous analysis~\cite{CJ14,CJ15,CJ17}, however, we found that the correspondence relation including only LF vertex functions given by Eq.~(\ref{eq30c})
brings about the self-consistency problem, i..e. the same physical quantity obtained from different components of the current and/or the polarization vectors
yields different results in the standard LFQM.
Furthermore, we also discovered the additional requirement for
the correct correspondence relation between the two models
to obtain the current-component independent physical observables in the standard LFQM.

Our new correspondence relation (denoted by  ``CJ-scheme" for convenience)
 to restore the self-consistency in the standard LFQM 
is given by~\cite{CJ14,CJ15,CJ17,Choi21}:
\be\label{eq31c}
\sqrt{2N_c} \frac{ \chi_{1(2)}(x,{\bf k}^{(\prime)}_\perp) } {1-x}
 \to \frac{ \phi_{1(2)}(x,{\bf k}^{(\prime)}_\perp) } {\sqrt{ {\cal A}^2_{1(2)} + {\bf k}^{(\prime)2}_\perp }}, \;\; M_{1(2)}\to M^{(\prime)}_0,
 \ee
that is, the physical mass $M_{1(2)}$ included in the integrand of the BS amplitude, 
e.g. the operators $\la{\cal O}\ra^{\rm BS}_{\rm on}$ in Eqs.~(31)-(35)
should be replaced by the invariant mass $M^{(\prime)}_0$ as all constituent quark and antiquarks are required to
be on  their respective mass shell in the standard LFQM.
We should note that this ``CJ-scheme" has been
verified through our previous analyses for the decay constants 
and the twist-2 and-3 DAs of pseudoscalar and vector mesons~\cite{CJ14,CJ15,CJ17}
and the form factor $f_-(q^2)$ for the semileptonic $B$ decays~\cite{Choi21}. 

We now show in this work that the ``CJ-scheme" is also valid to obtain the current-component independent tensor form factor $s(q^2)$ 
in addition to $f_-(q^2)$~\cite{Choi21}. 
That is, applying Eq.~(\ref{eq31c}) 
to the form factors ${\cal F}=\{f^{(+-)}_-, s^{(+-)} \}$ defined in 
 Eq.~(\ref{eq21c}) implies the replacements of the current operators
 $\la {\cal O}\ra_{\rm on}^{\rm BS}=\{ {\cal O}^{(+-)}_-, {\cal O}^{(+-)}_s \}$ in the BS model with 
 $\la {\cal O}\ra_{\rm on}^{\rm SLF}=\{ {\tilde {\cal O}}^{(+-)}_-, {\tilde {\cal O}}^{(+-)}_s \}$ in the standard LFQM, i.e.,
 \bea\label{eq32c}
{\tilde {\cal O}}^{(+-)}_- &=& -  \frac{S^+_{\rm on}}{2P^+_1} \biggl( \frac{\Delta M^2_{0+}  + {\bf q}^2_\perp}{\Delta M^2_{0-}  - {\bf q}^2_\perp} \biggr)
  + \frac{ P^+_1 S^-_{\rm on}}{ \Delta M^2_{0-}  - {\bf q}^2_\perp},\\
% \nonumber\\
{\tilde  {\cal O}}^{(+-)}_s &=& - \frac{i T^{+-}_{\rm on} }{2 (\Delta M^2_{0-}  - {\bf q}^2_\perp)},
\eea
where ${\Delta M}^2_{0\pm} = M^2_0 \pm M^{\prime 2}_0$.
Then, we obtain from Eqs.~(\ref{eq29c} ),~(\ref{eq32c}), and (46)
 the current-component independent form factors,  i.e.
$[f^{(+\perp)}_-]^{\rm SLF}_{\rm on} \doteq [f^{(+-)}_-]^{\rm SLF}_{\rm on}$ and
$[s^{(+-)}]^{\rm SLF}_{\rm on} \doteq [s^{(+\perp)}]^{\rm SLF}_{\rm on}$ in the standard LFQM, 
where ``$\doteq$" represents the equality of both sides numerically. 
The additional requirement in the ``CJ-scheme",  i.e. $M_{1(2)}\to M^{(\prime)}_0$,  can therefore be
regarded as the effective inclusion of the zero modes in the valence region of the $q^+=0$ frame in the standard LFQM.
This replacement $M_{1(2)}\to M^{(\prime)}_0$
is not possible in the BS model due to the form of the LF vertex function $\chi$ given by Eq.~(\ref{eq17c}).
 
The final results for $f_+$, $f_-= f^{(+\perp)}_-\doteq f^{(+-)}_-$,  and $s=s^{(+\perp)}\doteq s^{(+-)}$ in the standard LFQM 
are given by
\begin{widetext}
\bea\label{eq33c}
f_+ (q^2) &=& \int^{1}_{0}dx\int \frac{d^{2}{\bf k}_{\perp}}{16\pi^3}
\frac{\phi_{1}(x,{\bf k}_{\perp})}{\sqrt{ {\cal A}_{1}^{2} + {\bf k}^{2}_{\perp}}}
\frac{\phi_{2}(x,{\bf k}'_{\perp})}{\sqrt{ {\cal A}_{2}^{2}+ {\bf k}^{\prime 2}_{\perp}}}
%\nonumber\\
%&&\times 
( {\cal A}_{1}{\cal A}_{2}+{\bf k}_{\perp}\cdot{\bf k'}_{\perp} ),
\eea
%\nonumber\\
%\eea
%\bea\label{eq34c}
\bea
 f^{(+\perp)}_- (q^2)&=& \int^1_0 {\bar x} dx
  \int \frac{ d^2{\bf k}_\perp } { 16\pi^3 }
  \frac{ \phi_1 (x, {\bf k}_\perp) } {\sqrt{ {\cal A}^2_1 + {\bf k}^2_\perp }}
  \frac{ \phi_2 (x, {\bf k'}_\perp) } {\sqrt{ {\cal A}^2_2 + {\bf k}^{\prime 2}_\perp }}
  \nonumber\\
 &&\times
  \biggl\{ - {\bar x} M^2_0 + (m_2 - m_q){\cal A}_1 - m_q (m_1 - m_q)
%  \nonumber\\
%  &&
  + \frac{ {\bf k}_\perp\cdot{\bf q}_\perp }{q^2} [ M^2_0 + M'^2_0
  - 2(m_1 - m_q) (m_2 - m_q) ]
  \biggr\},
%  \nonumber\\
\eea
\bea
%   and 
%  \bea\label{eq35c}
f^{(+-)}_{-}(q^2) &=&
 \int^1_0\frac{dx}{x^2} \int\frac{d^2{\bf k}_\perp}{16\pi^3} 
 \frac{\phi_{1}(x,{\bf k}_{\perp})}{\sqrt{ {\cal A}_{1}^{2} + {\bf k}^{2}_{\perp}}}
\frac{\phi_{2}(x,{\bf k}'_{\perp})}{\sqrt{ {\cal A}_{2}^{2}+ {\bf k}^{\prime 2}_{\perp}}}
\nonumber\\
&&\times
\biggl\{ 
a_0
  \biggl[ x^2 {\bar x} M^2_0 ({\bf k}^2_\perp + {\bf k}_\perp\cdot{\bf q}_\perp)
%  \nonumber\\
%  &&
  + {\bar x} (m_1 {\cal A}_1 + {\bf k}^2_\perp) [m^2_2 + ({\bf k}_\perp + {\bf q}_\perp)^2]
%\nonumber\\
%&&
+ x^2 m_1 m_2 (m^2_{q} + {\bf k}^2_\perp) 
%\nonumber\\
%&& 
+ x {\bar x}  m_2 m_{q} (m^2_1 + {\bf k}^2_\perp)
\biggr]
\nonumber\\
&&~~~~~
- x^2  b_0 ( {\bf k}_\perp\cdot{\bf k'}_\perp + {\cal A}_1{\cal A}_2 )
\biggr\},
 \eea
 for the vector current, and 
% \end{widetext}
%\end{widetext}
%\begin{widetext}
\bea\label{eq34c}
s^{(+\perp)}(q^2) &=& - \int^{1}_{0} (1-x) dx \int \frac{d^{2}{\bf k}_{\perp}}{16\pi^3}
\frac{\phi_{1}(x,{\bf k}_{\perp})}{\sqrt{ {\cal A}_{1}^{2} + {\bf k}^{2}_{\perp}}}
\frac{\phi_{2}(x,{\bf k}'_{\perp})}{\sqrt{ {\cal A}_{2}^{2}+ {\bf k}^{\prime 2}_{\perp}}} 
 \left[  (m_1 - m_2) \frac{  {\bf k}_\perp\cdot{\bf q}_\perp }{{\bf q}_\perp^2} + {\cal A}_1 \right],
 \\
% \nonumber\\
%\eea
%and
% \bea\label{ap:7}
s^{(+-)}(q^2)&=& \int^{1}_{0} \frac{dx}{x} \int \frac{d^{2}{\bf k}_{\perp}}{16\pi^3}
\frac{\phi_{1}(x,{\bf k}_{\perp})}{\sqrt{ {\cal A}_{1}^{2} + {\bf k}^{2}_{\perp}}}
\frac{\phi_{2}(x,{\bf k}'_{\perp})}{\sqrt{ {\cal A}_{2}^{2}+ {\bf k}^{\prime 2}_{\perp}}}
 \nonumber\\
 &&\times
  \frac{a_0}{2}
 \biggl[ (1- 2x) (m_1 - m_2) {\bf k}^2_\perp + 2 (1-x) {\cal A}_1 {\bf k}_\perp\cdot{\bf q}_\perp
+ (1-x) {\cal A}_1 {\bf q}^2_\perp + (m_2 - m_1) {\cal A}_1 {\cal A}_2 
 \biggr],
\eea
\end{widetext}
for the tensor current, where $a_0 = \frac{2}{M^2_0 - M'^2_0 -{\bf q}_\perp^2}$
and $b_0 = \frac{M^2_0 + M'^2_0 + {\bf q}_\perp^2}{M^2_0 - M'^2_0 -{\bf q}_\perp^2}$. 
Indeed, our prescription $M_{1(2)}\to M^{(\prime)}_0$ is applied through the two terms $(a_0, b_0)$ in 
$f^{(+-)}_-$ and $s^{(+-)}$.
Finally, we confirm from the numerical calculations the current independencies of the form factors, 
i.e. $f_-(q^2)=f^{(+\perp)}_-\doteq f^{(+-)}_-$ 
and $s(q^2)=s^{(+\perp)}\doteq s^{(+-)}$, which supports the universality of the ``CJ-scheme"
given by Eq.~(\ref{eq31c}) 
and the self-consistency of our standard LFQM.

For our numerical calculations in the following section,
we use the tensor form factor $f_T(q^2)=s(q^2) (M_1 + M_2)$ as defined in~Eq.~(\ref{eq4c}). 
We should emphasize that the physical masses $M_{1(2)}$ used in defining $f_T$ is nothing to do with our
correspondence relations. Only the physical masses $M_{1(2)}$  appeared as a result from the choice of  minus component ($\mu, \nu=-$) 
of the vector and tensor currents given by Eqs.~(\ref{eq1c}) and~(\ref{eq2c}) are eligible for the transformation
into the corresponding invariant masses $M^{(\prime)}_0$ as shown in Eq.~(10).

\section{Numerical Results}
\label{sec:IV}
In our numerical calculations for the semileptonic and rare
$D\to (\pi, K)$ decays, we use the
model parameters ($m_{q\bar{q}},\beta_{q{\bar q}}$) for the harmonic oscillator (HO)
confining potential given in Table~\ref{t1} obtained from the calculation of the ground state
meson mass spectra~\cite{Choi07,CJ09}. The decay constants of $(\pi, K, D)$ mesons obtained 
from the HO parameters are given by
$(f_\pi, f_K, f_D) = (131, 155, 197)$ MeV compared to the experimental data~\cite{PDG},
$(f^{\rm exp.}_\pi, f^{\rm exp.}_K, f^{\rm exp.}_D) = (130.2 (1.2), 155.7(3), 212.6(7))$ MeV.
While the decay constant of $D$ meson is not quite sensitive to the quark mass variation, e.g.
$f_D = 199^{-2}_{+1}$ MeV for $m_c = 1.7^{+0.1}_{-0.1}$ GeV, we find that the form factors
are somewhat sensitive to $m_c$. Thus, as a sensitivity check of our LFQM, 
we use this charm quark mass variation for the calculations of the form factors and
the branching ratios. 
For the physical $(D, K, \pi)$ meson masses, we
use the central values quoted by the Particle Data Group (PDG)~\cite{PDG}.

\begin{table}[t]
\caption{The constituent quark mass [GeV] and the Gaussian parameters
$\beta$ [GeV] for the HO potential obtained by the variational
principle~\cite{CJ09,Choi07}. $q=u$ and $d$.}\label{t1}
\renewcommand{\arraystretch}{1.2}
\setlength{\tabcolsep}{5pt}
\begin{tabular}{cccccc} \hline\hline
 $m_q$ & $m_s$ & $m_c$ & $\beta_{qq}$ & $\beta_{qc}$ & $\beta_{sc}$  \\
\hline
0.25 & 0.48 & 1.8 & 0.3194 & 0.4216 & 0.4686    \\
\hline\hline
\end{tabular}
\end{table}

In principle, it is possible to use the $q^+\neq 0$ frame satisfying  $q^2 = q^+ q^-  -{\bf q}^2_\perp >0$ for this timelike semileptonic and rare decays.
However, in this $q^+\neq 0$ frame, 
it  is  inevitable  to  confront  the particle-number-nonconserving Fock state (or nonvalence) contribution~\cite{CJ01,BH98}.
The main source of difficulty in the LFQM phenomenology is the  paucity of information on
the non-wave-function vertex~\cite{BCJ2} in the nonvalence diagram arising from the quark-antiquark pair creation/annihilation. 
This should contrast with the usual  LFWF used in the valence region. 
Contrary to the $q^+\neq 0$ frame, the $q^+=0$ frame does not suffer from the nonvalence contribution although one needs to be
cautious about the zero-mode problem as we discussed already. Once the zero-mode issue is resolved as we proved in this work,
it is straightforward to analytically continue the form factors  given by Eqs.~(47)-(51) 
obtained in the spacelike region to the timelike physical region.

%This should contrast with the usual  LFWF in the valence region. In principle, there is a systematic program
%as was discussed in~\cite{BH98} to include the particle-number-nonconserving amplitude to take into
%account the nonvalence contributions. However, the program requires to find all the higher Fock-state wave
%functions while there has been relatively little progress in computing the basic wave functions of hadrons from
%first principles. 

Our results of the form factors  $(f_\pm, f_0, f_T)$ obtained from Eqs.~(47)-(51)   can also
be compared with several parametric forms. Among several forms, a more systematic and model-independent
parametrization of semileptonic form factors, 
often referred to as ``$z$-expansion" or ``$z$-parametrization"~\cite{z1,z2},
 has been developed based on  general properties of analyticity, unitarity,  and crossing symmetries. Especially, this 
 $z$-parametrization provides better control of theoretical uncertainties in LQCD calculations~\cite{Lub17,Lub18} .

Our direct LFQM results for the form factors $f_i (q^2)$ $(i=\pm, 0, T)$
are also well described by the ``$z$-parametrization", which takes the form~\cite{Lub17,Lub18}
\be\label{eq35c}
f_j(q^2) = \frac{ f_j(0) + c_j (z - z_0)\left(1 + \frac{z+z_0}{2}\right)}{1 - b_j q^2},
\ee
where 
\be\label{eq36c}
z= \frac{ \sqrt{t_+ - q^2} - \sqrt{t_+ - t_0}} { \sqrt{t_+ - q^2} + \sqrt{t_+ - t_0}},
\ee
and $z_0=z(q^2=0)$ with $t_\pm = (M_1 \pm M_2)^2$ and $t_0 = t_+ (1 - \sqrt{1 - t_-/t_+})$.

The fitted parameters $(b_j, c_j) (j=+, 0, T)$ for the $D\to\pi$ and $D\to K$ TFFs $(f_+, f_0, f_T)$ are summarized 
in Tables~\ref{t2} and~\ref{t3}, respectively, where
the errors occur due to the choice of $m_c =1.7^{+0.1}_{-0.1}$ GeV. 
In Table~\ref{t4}, we also compare the form factors $f_+(0)$ and $|f_T(0)|$ for $D\to(\pi, K)$ transitions at $q^2=0$ with those obtained 
from various theoretical model predictions and experimental data~\cite{BES15, BABAR15,BABAR07}.
%We should note that our results for the current independences, i.e. $f_-(q^2)=f^{(+\perp)}_-=f^{(+-)}_-$ and $s(q^2)=s^{(+\perp)}=s^{(+-)}$,  
%are also  confirmed numerically in this parametric formulations.
%This substantiates the self-consistency of our LFQM.

\begin{table*}
%[htbp]
\caption{Fitted parameters $(b_j, c_j)$ in Eq.~(\ref{eq35c}) for the $D\to\pi$ TFFs with $m_c=1.7^{+0.1}_{-0.1}$ GeV.  $b_j$ is in unit of  [GeV$^{-2}$]. }
\label{t2}
\renewcommand{\arraystretch}{1.2}
\setlength{\tabcolsep}{2.2pt}
\centering
\begin{tabular}{cccccccc} \hline\hline
$f_{(+,0)}(0)$ &  $b_+$  & $c_+$ &  $b_0$ & $c_0$ & $f_{T}(0)$ &  $b_T$ & $c_T$  \\
\hline
$0.613^{-(21)}_{+(22)}$ &  $0.1899^{-(208)}_{+(233)}$ & $-0.8200^{+(275)}_{-(317)}$ &
$0.2986^{-0.0882}_{+11.8745}$ &  $1.9051^{-0.5986}_{+107.077}$ & 
$-0.501^{+(36)}_{-(39)}$ &  $0.1957^{-(216)}_{+(242)}$ & $0.6290^{-(417)}_{+(481)}$ \\
% \hline
%ETMC~\cite{Lub17,Lub18}&  0.612(35) & 0.1314(127) & $-1.985(347)$ & 0.0342(122) & $-1.188(256)$ & $-0.506(79)$ & 0.1461(681) & 1.10 (1.03)
% \\
\hline\hline
\end{tabular}
%$^{*}$ We used our notation, i.e. the negative values of $f_T(q^2)$.
\end{table*}

\begin{table*}
%[htbp]
\caption{Fitted parameters $(b_j, c_j)$ in Eq.~(\ref{eq35c}) for the $D\to K$ decay with $m_c=1.7^{+0.1}_{-0.1}$ GeV. 
$b_j$ is in unit of  [GeV$^{-2}$].}
\label{t3}
\renewcommand{\arraystretch}{1.2}
\setlength{\tabcolsep}{2.2pt}
\centering
\begin{tabular}{cccccccc} \hline\hline
$f_{(+,0)}(0)$ &  $b_+$  & $c_+$ &  $b_0$   & $c_0$ & $f_{T}(0)$ &  $b_T$  & $c_T$  \\
\hline
$0.744^{-(22)}_{+(23)}$ &  $0.1787^{-(157)}_{+(176)}$ & $-1.1711^{+(571)}_{-(618)}$ &
$-0.0563^{+5.0348}_{+0.1426}$ &  $-2.3039^{+78.3509}_{+1.8289}$ & 
$-0.660^{+(42)}_{-(45)}$ &  $0.1826^{-(163)}_{+(182)}$ & $0.9893^{-(794)}_{+(897)}$\\
%\hline
%ETMC~\cite{Lub17,Lub18} & 0.764(31) & $M^{-2}_{D^*_s}$ & $-0.066(333)$ & $M^{-2}_{D^*_s}$ & $-2.084(283)$ & $-0.687(54)$ & 0.0854(671) &  2.86 (1.46) \\
\hline\hline
\end{tabular}
\end{table*}

%%%%% Table 3
\begin{table*}
%[t]
\caption{Form factors $f_+(0)$ and $|f_T(0)|$ for $D\to(\pi, K)$ transitions at $q^2=0$ compared with various model predictions and experimental data.}\label{t4}
\renewcommand{\arraystretch}{1.2}
\setlength{\tabcolsep}{5pt}
\begin{tabular}{cccccccccc} \hline\hline
$F(0)$ & This work  & ~\cite{Lub17,Lub18} &~\cite{Ball06}  &~\cite{FGK20} & ~\cite{Ivanov19} &~\cite{QC20}   &~\cite{Verma12} &~BES III~\cite{BES15} & BABAR~\cite{BABAR15,BABAR07}\\
\hline
 $f^{D\pi}_+(0)$ & $0.613^{-(21)}_{+(22)}$  & 0.612 (35) & 0.63 (11) & 0.640 & 0.63 (9)  &$-$  &  0.66 (1) & 0.637 (24) & 0.610 (25) \\
 \hline
$|f^{D\pi}_T(0)|$ & $0.501^{+(36)}_{-(39)}$ & 0.506 (79) & $-$ &$-$ & $-$ & $0.84^{+(16)}_{-(13)}$  & $-$ & $-$ & $-$  \\
 \hline
$f^{DK}_+(0)$ & $0.744^{-(22)}_{+(23)}$  & 0.765 (31) & 0.75 (12) & 0.716 & 0.77 (11) &  $-$  & 0.79 (1) & 0.737 (4) & 0.727 (11) \\
 \hline
$|f^{DK}_T(0)|$ & $0.660^{+(42)}_{-(45)}$ & 0.687 (54)& $-$  & $-$ & $-$ & $0.96^{+(17)}_{-(15)}$ & $-$  & $-$ & $-$\\ 
\hline\hline
\end{tabular}
\end{table*}
%%%%%%%%

\begin{figure}[htbp]
\centering
\includegraphics[height=7cm, width=7cm]{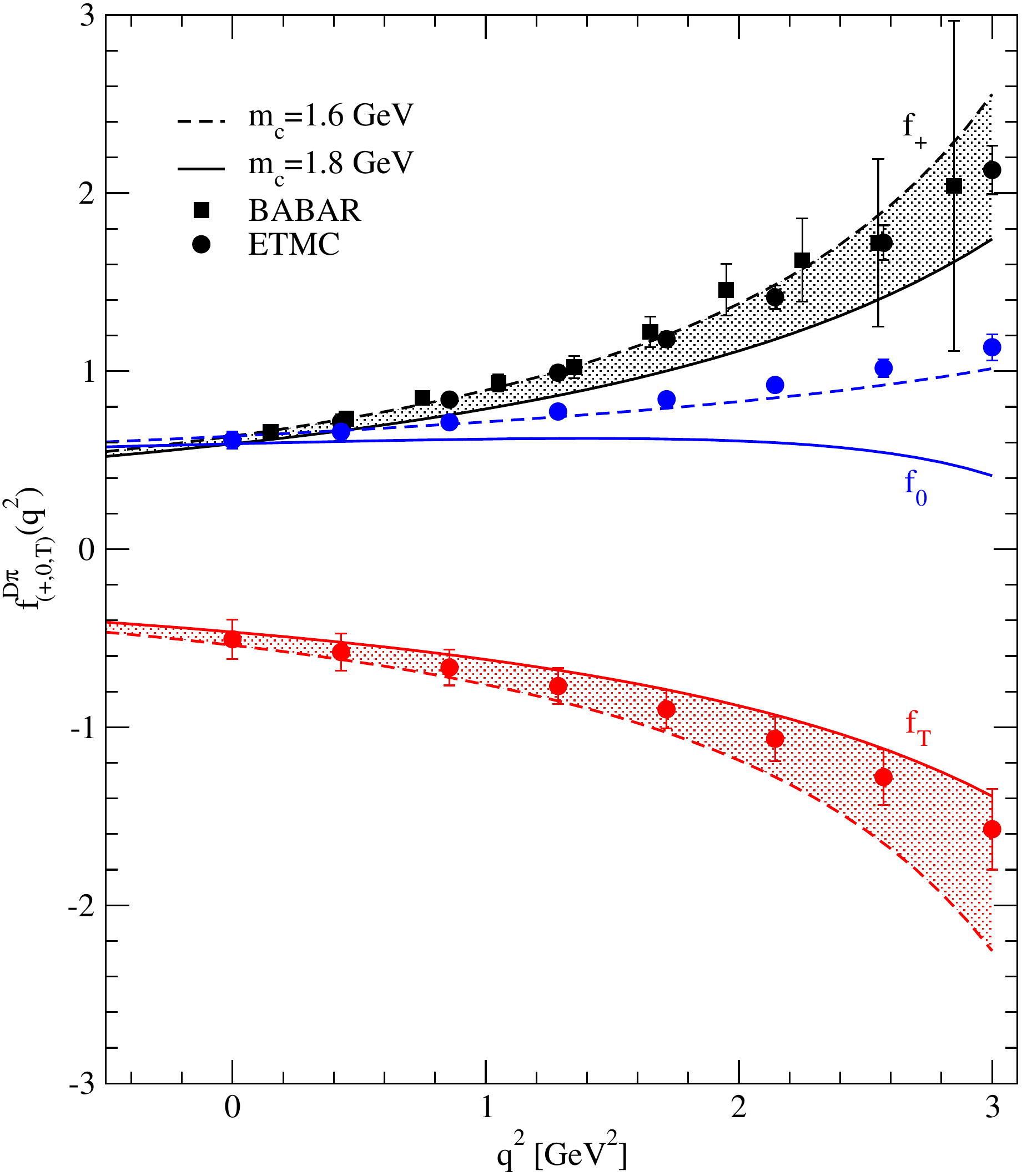}
\caption{\label{fig1} (Color online): The $q^2$ dependent form factors ($f_+, f_0, f_T$) of the $D\to\pi$ decay
for both spacelike and the kinematic timelike regions, $-0.5\leq q^2\leq (M_D-M_\pi)^2$ GeV$^2$. 
For comparison, the data taken from the  LQCD (circles)~\cite{Lub17,Lub18} and $BABAR$~\cite{BABAR15}  (squares) 
are shown. }
\end{figure}

In Fig.~\ref{fig1}, we show the $q^2$ dependences of $f^{D\pi}_+(q^2)$ (black lines), $f^{D\pi}_0(q^2)$ (blue lines),
and $f^{D\pi}_T(q^2)$ (red lines) for $D\to \pi$ decay, where  the solid and dashed lines represent the results obtained from
$m_c=1.8$ GeV and 1.6 GeV, respectively. That is, the bands correspond to the sensitivity coming from the charm quark 
mass variation, $m_c=1.7^{+0.1}_{-0.1}$ GeV in our LFQM. 
We should note that the form factors are displayed not only for 
the whole timelike kinematic region [$0 \leq q^2\leq (M_D - M_\pi)^2$] (in unit of GeV$^2$)
but also for the spacelike region ($-0.5\leq q^2\leq 0$) (in unit of GeV$^2$) to demonstrate the validity of our analytic continuation from
spacelike region to the timelike one by changing ${\bf q}^2_\perp$ to $-q^2$ in the form factors.
For comparison, the data (circles) of the form factors $(f_+, f_0, f_T)$  from the LQCD (for the ETM Collaboration)~\cite{Lub17,Lub18} and
the data of $f_+$ (squares) extracted from the $BABAR$~\cite{BABAR15} are shown. 
Our results for  $f^{D\pi}_+(0)=0.613^{-(21)}_{+(22)}$ and  $|f^{D\pi}_T(0)|=0.501^{+(36)}_{-(39)}$
are  in good agreement with
$f^{D\pi}_+(0)=0.610(25)$ from the $BABAR$~\cite{BABAR15} and $f^{D\pi}_+(0)=0.637(24)$ from  the BES III~\cite{BES15},
as well as $f^{D\pi}_+(0)=0.612(35)$ and $|f^{D\pi}_T(0)|=0.506(79)$ from the LQCD~\cite{Lub17,Lub18}.
As one can see from Fig.~\ref{fig1},  the sensitivity to the charm quark mass is more pronounced at  the zero-recoil ($q^2=q^2_{\rm max}$) of the
final meson than the maximum recoil ($q^2=0$).
Especially,  the $q^2$-dependent behaviors of our results
show better agreement with the data from the $BABAR$~\cite{BABAR15} and LQCD~\cite{Lub17,Lub18} when we use
$m_c\simeq 1.6$ GeV rather than 1.8 GeV.

\begin{figure}[htbp]
\centering
\includegraphics[height=7cm, width=7cm]{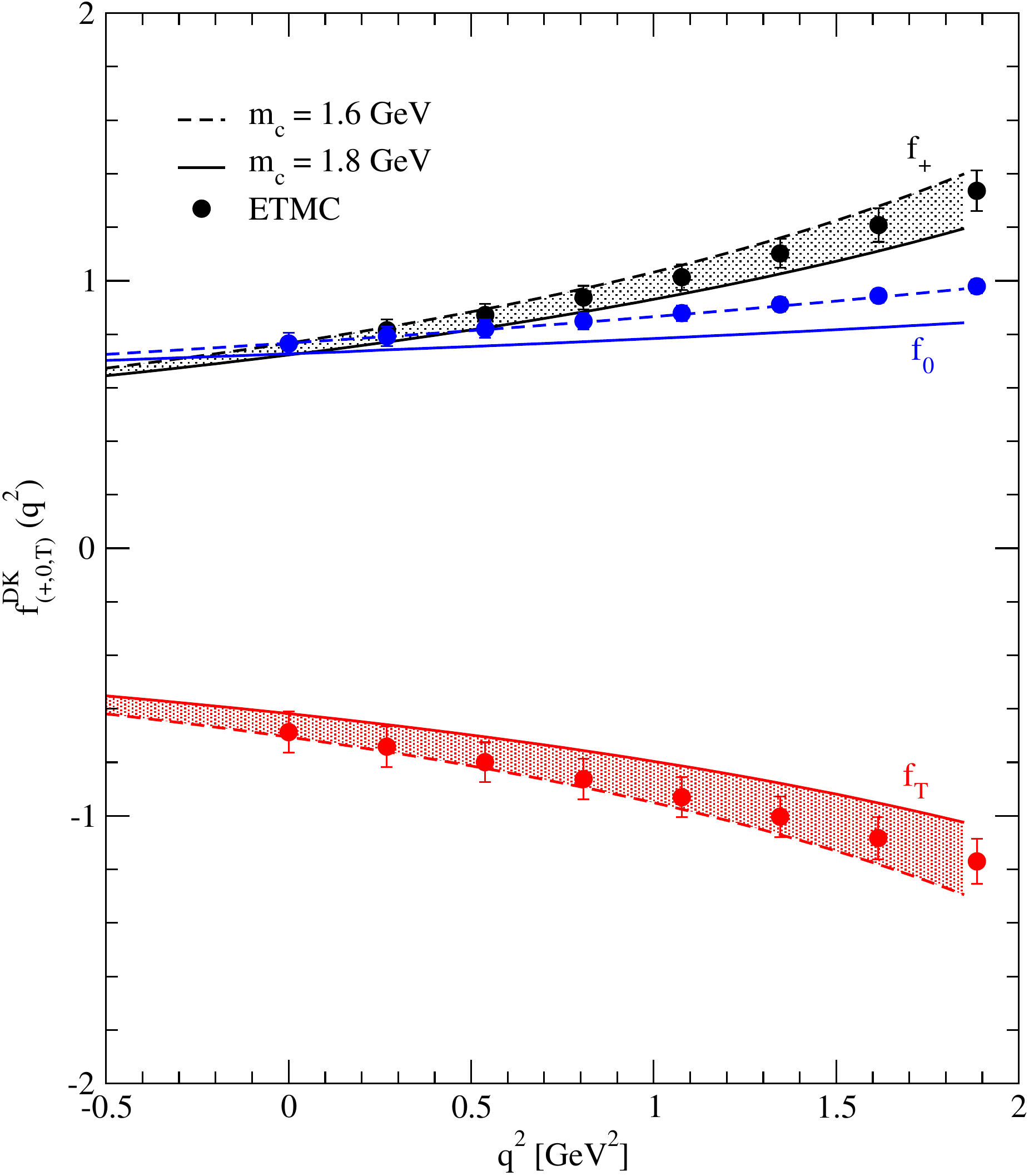}
\caption{\label{fig2} (Color online): The $q^2$ dependent form factors ($f_+, f_0, f_T$) of the $D\to K$ decay
for both spacelike and the kinematic timelike regions, $-0.5\leq q^2\leq (M_D-M_K)^2$ GeV$^2$.
For comparison, the data taken from the LQCD (circles)~\cite{Lub17,Lub18} are shown.}
\end{figure}

In Fig.~\ref{fig2}, we show the $q^2$ dependences of $(f^{DK}_+, f^{DK}_0, f^{DK}_T)$ for $D\to K$ decay, compared with
the results from the LQCD~\cite{Lub17,Lub18}. The same line codes are used as in Fig.~\ref{fig1}. \
Our predictions of $f^{DK}_+(0)=0.744^{-(22)}_{+(23)}$ and $|f^{DK}_T(0)|=0.660^{+(42)}_{-(45)}$
agree with $f^{DK}_+(0)=0.737(4)$ from the BES III~\cite{BES15} and
 $f^{DK}_+(0)=0.727(11)$ from the $BABAR$~\cite{BABAR07},
as well as 
$f^{DK}_+(0)=0.764(31)$ and  $|f^{DK}_T(0)|=0.687(54)$ from  the LQCD~\cite{Lub17,Lub18} within the error bars. 
As in the case of $D\to\pi$ decay,  the $q^2$-dependent behaviors of our results
show better agreement with the data from the LQCD~\cite{Lub17,Lub18} when we use
$m_c\simeq 1.6$ GeV rather than 1.8 GeV.

\begin{figure}
%[htbp]
\centering
\includegraphics[height=7cm, width=7cm]{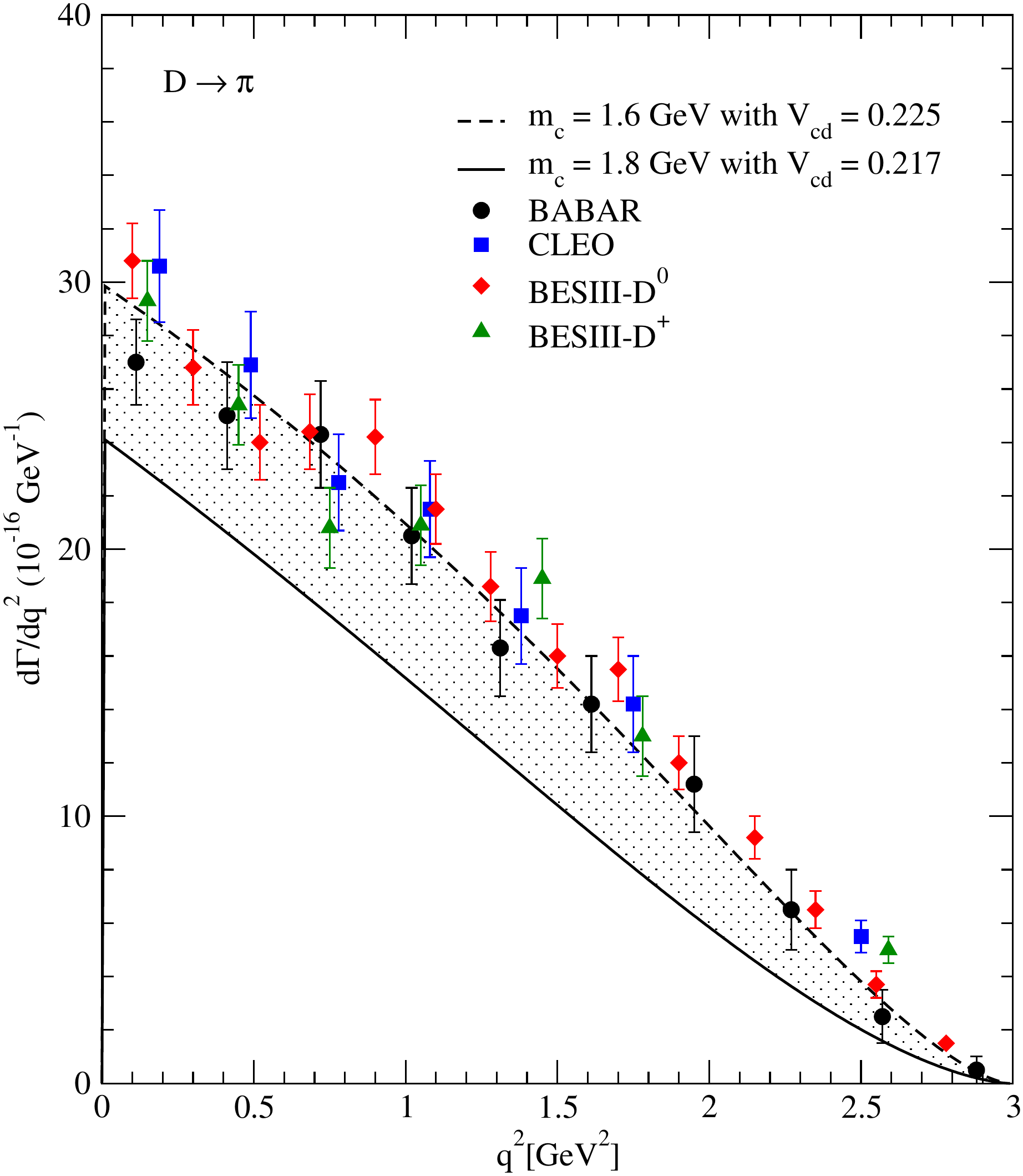}
\caption{\label{fig3} Differential decay rate for the $D\to \pi e\nu_e$ decay compared with the
 experimental data from the $BABAR$~\cite{BABAR15,BABAR07} (black circles), CLEO~\cite{CLEO09} (blue squares),
and BES III~\cite{BES15,BES1} for neutral $D^0$ (red diamonds) 
and charged $D^+$ with the account of isospin factor (green triangles).}
\end{figure}
\begin{figure}
%[htbp]
\centering
\includegraphics[height=7cm, width=7cm]{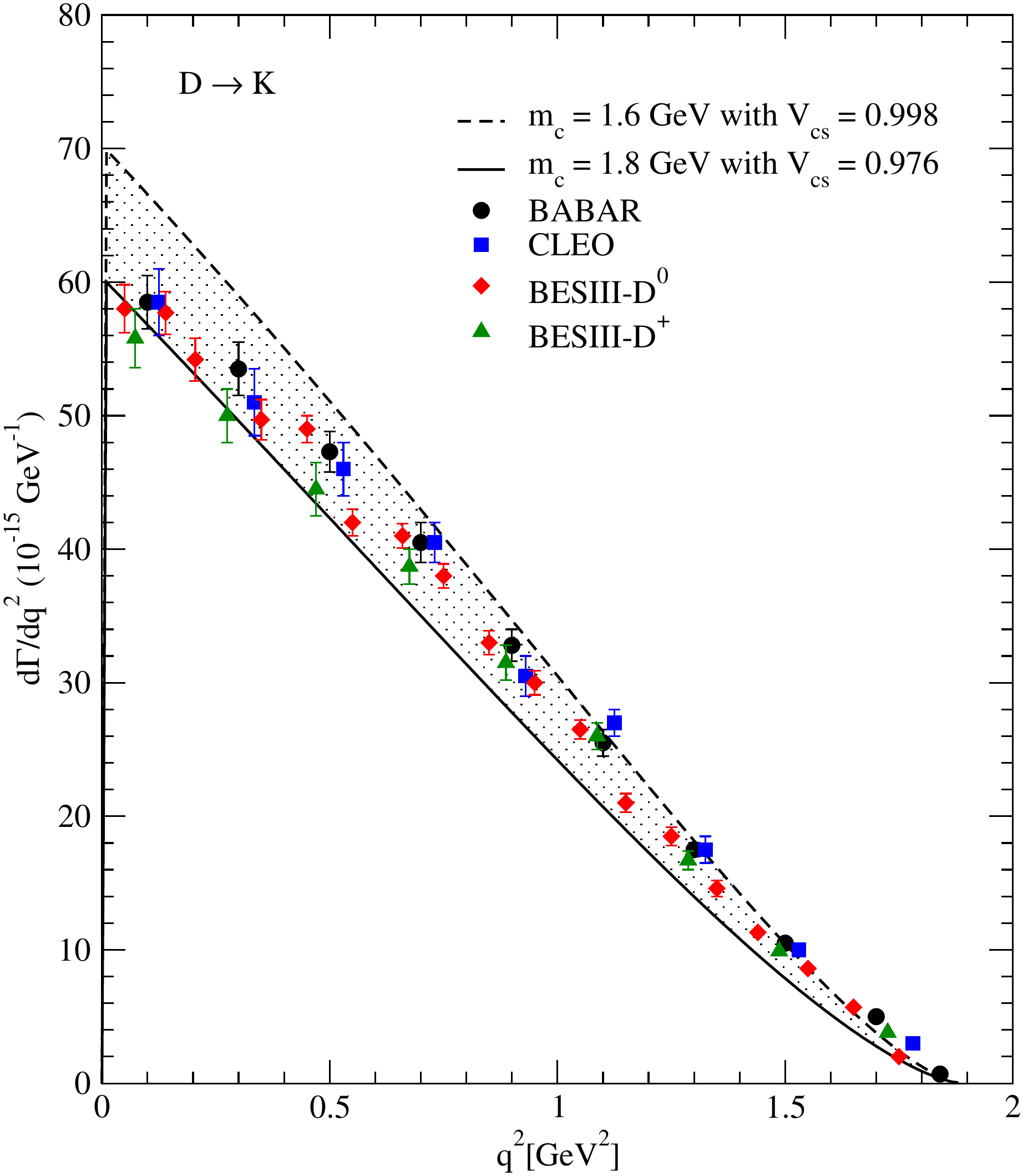}
\caption{\label{fig4}Differential decay rate for the $D\to K e\nu_e$ decay compared with the
 experimental data from the $BABAR$~\cite{BABAR15,BABAR07} (black circles), CLEO~\cite{CLEO09} (blue squares),
and BES III~\cite{BES15,BES1} for neutral $D^0$ (red diamonds) 
and charged $D^+$ with the account of isospin factor (green triangles).}
\end{figure}

Figs.~\ref{fig3} and~\ref{fig4} 
show our predictions for the differential decay rates  of $D\to \pi e\nu_e$ and
$D\to K e\nu_e$ decays, respectively, compared with the
experimental data from the $BABAR$~\cite{BABAR15,BABAR07} (black circles), CLEO~\cite{CLEO09} (blue squares),
and BES III~\cite{BES15,BES1} for neutral $D^0$ (red diamonds) 
and charged $D^+$ with the account of isospin factor (green triangles).
In our numerical calculations of the branching ratios,
we use the CKM matrix elements $|V_{cd}|=0.221\pm 0.004$ and
$|V_{cs}|=0.987\pm 0.011$ quoted by the PDG~\cite{PDG}.
Considering uncertainties coming from  the CKM elements and the constituent charm quark mass
$m_c=1.7^{+0.1}_{-0.1}$ GeV, we made band plots, i.e.
the solid (dashed) lines represent the results obtained from $m_c=1.8 (1.6)$ GeV with
lower (upper) limits of the CKM elements.
Our results are shown to be consistent with the current available experimental data within those uncertainties.
%%%%%%%Table

\begin{table}
%[htbp]
\caption{Branching ratios (in $10^{-3}$) for $D\to \pi\ell\nu_\ell$ ($\ell=e,\mu$) obtained from
using $m_c=1.7^{+0.1}_{-0.1}$ GeV together with $|V_{cd}|=0.221\pm 0.004$~\cite{PDG}.}
\label{t5}
\renewcommand{\arraystretch}{1.2}
\setlength{\tabcolsep}{12pt}
\centering
\begin{tabular}{ccc} \hline\hline
 Channel &  Ours & PDG~\cite{PDG}  \\
\hline
$D^+\to \pi^0 e^+\nu_{e}$ & $3.03^{+0.57}_{-0.44}$ & $3.72\pm 0.17$ \\
$D^+\to \pi^0 \mu^+\nu_{\mu}$ & $2.97^{+0.58}_{-0.44}$ & $3.50\pm 0.15$ \\
$D^0\to \pi^- e^+\nu_{e}$ & $2.37^{+0.45}_{-0.34}$ & $2.91\pm 0.04$ \\
$D^0\to \pi^- \mu^+\nu_{\mu}$ & $2.33^{+0.44}_{-0.34}$ & $2.67\pm 0.12$ \\
\hline\hline
\end{tabular}
\end{table}
\begin{table}
%[htbp]
\caption{Branching ratios (in $\%$) for $D\to K\ell\nu_\ell$ ($\ell=e,\mu$) decays obtained from
using $m_c=1.7^{+0.1}_{-0.1}$ GeV together with $|V_{cs}|=0.987\pm 0.011$~\cite{PDG}.}
\label{t6}
\renewcommand{\arraystretch}{1.2}
\setlength{\tabcolsep}{12pt}
\centering
\begin{tabular}{ccc} \hline\hline
 Channel &  Ours & PDG~\cite{PDG}  \\
\hline
$D^+\to {\bar K}^0 e^+\nu_{e}$ & $8.88^{+1.10}_{-0.80}$ & $8.73\pm 0.10$ \\
$D^+\to {\bar K}^0 \mu^+\nu_{\mu}$ & $8.64^{+1.07}_{-0.79}$ & $8.76\pm 0.19$ \\
$D^0\to K^- e^+\nu_{e}$ & $3.50^{+0.43}_{-0.31}$ & $3.54\pm 0.034$ \\
$D^0\to K^- \mu^+\nu_{\mu}$ & $3.41^{+0.41}_{-0.30}$ & $3.41\pm 0.004$ \\
\hline\hline1
\end{tabular}
\end{table}

In Tables~\ref{t5} and~\ref{t6}, we summarize our results for the branching 
ratios  for $D\to \pi\ell\nu_\ell$ and $D\to K\ell\nu_\ell$ ($\ell=e,\mu$), respectively,
and compare with the experimental data from PDG~\cite{PDG}.
Our results for Br($D\to\pi$) are best fit to the data with $m_c=1.6$ GeV but those for Br($D\to K$) 
prefers $m_c=1.7$ GeV.

Finally, as a test for the LFU,  the $R$ ratios of the semileptonic $D\to (\pi, K)$ decays
is defined by
\be
R_P = \frac{{\rm Br}(D\to P\mu\nu_\mu)}{{\rm Br}(D\to P e\nu_e)},
\ee
where $P = (\pi, K)$. 
Our predictions for  $R_P$ obtained from using $m_c=1.7^{+0.1}_{-0.1}$ GeV are as follows:
$(R_{\pi^0},  R_{\pi^-}) = (0.980^{+0.165}_{-0.003}, 0.983^{-0.001}_{-0.003})$,
$(R_{K^-}, R_{{\bar K}^0}) = (0.974^{-0.002}_{+0.001}, 0.973^{-0.001}_{-0.001})$.
Our results are consistent with the recent measurements from the BES III,
$(R_{\pi^0}, R_{\pi^-}) = (0.942 \pm 0.046, 0.905\pm 0.035)$~\cite{BES2}, and
$R_{K^-}=0.974\pm 0.014$~\cite{BES5}, as well as other theoretical predictions such as
$R_{\pi}=0.985(2)$ and $R_{K}=0.975(1)$ from the LQCD~\cite{Lat18},
$R_{\pi}=0.985$ and $R_{K}=0.980$ from the RQM~\cite{FGK20}, and
$R_{\pi}=0.98$ and $R_{K}=0.97$ from the CCQM~\cite{Ivanov19}.

%%%%%~

\section{Summary and Discussion}
\label{sec:V}
In this work, we discussed the self-consistence description on  the weak
form factors $f_+$, $f_-$ (or $f_0$), and $f_T$ for the exclusive
semileptonic  $D\to (\pi,K)\ell\nu_\ell$$(\ell=e, \mu,\tau)$ and rare $D\to (\pi,K)\ell\ell$ decays in the standard
LFQM. It has been known in the LF formulation that while the plus component ($J^+$) of the LF current $J^\mu$ in 
the matrix element can be regarded as  the ``good" current, the perpendicular (${\bf J}_\perp$) and
the minus ($J^-$) components of the current were known as the ``bad" currents since $({\bf J}_\perp, J^-)$
are easily contaminated by the treacherous points
such as the LF zero mode and the off-mass shell instantaneous contributions.

For a systematic analysis of such treacherous points in case one cannot avoid the use of ${\bf J}_\perp$ or $J^-$, 
we utilized the exactly solvable manifestly covariant BS model to carry out the LF calculations for three 
form factors, $(f_+, f_-, f_T)$. In particular, we obtained $f_-$ from two sets of the vector current,
$(J^+, {\bf J}_\perp)_V$ and $(J^+, J^-)_V$, and $f_T$ from two sets of tensor current, $J^{+\perp}_T$ and $J^{+-}_T$.
In this BS model, we found that while $f_+$ obtained from $J^+$ and $f_T$ obtained from $J^{+\perp}_T$ are
free from the zero modes, $f_-$ obtained from both $(J^+, {\bf J}_\perp)_V$ and $(J^+, J^-)_V$ sets and 
$f_T$ obtained from $J^{+-}_T$ receive the zero-mode contributions as well as the instantaneous ones.
We then linked the BS model to the standard LFQM 
using the ``CJ-scheme" ~\cite{CJ14,CJ15,CJ17,Choi21} given by Eq.~(\ref{eq31c}) for the correspondences
between the two models and replaced the LF vertex function in the BS model with the more
phenomenologically accessible Gaussian wave function provided by the LFQM analysis of meson mass spectra~\cite{CJ99,CJ99PLB}.
As in the case of previous analysis~\cite{CJ14,CJ15,CJ17,Choi21}, it is  astonishing to discover that the zero modes and the instantaneous contributions
present in the BS model become absent in the LFQM. In other words, 
our LFQM results of $(f_-,f_T)$ are shown to be independent of the components of the
current without involving any of those treacherous contributions.  Since the absence of the zero mode found in the standard LFQM
is mainly due to the replacement of the physical mass $M_{1(2)}$ with the invariant mass $M^{(\prime)}_0$ in the course of linking the two
models, this replacement could be regarded as an effective treatment of the zero mode in the standard LFQM. 

In the standard LFQM, the constituent quark and antiquark in a bound state are
required to be on-mass shell, which is different from the covariant formalism, in which the constituents are off-mass shell. 
The common feature of the standard LFQM is thus to use the sum of the LF energy of the constituent quark and antiquark for the meson mass
in the spin-orbit wave function, which is obtained by the interaction-independent Melosh transformation from the ordinary equal-time static
 spin-orbit wave function assigned by the quantum number $J^{PC}$. Under these circumstances, it is natural to apply the replacement 
 $M_{1(2)} \to M^{(\prime)}_0$ in the calculation of the physical observables in the standard LFQM. Indeed, we have shown explicitly that this correspondence 
 relation for the calculations of the decay constants and weak transition form factors between two pseudoscalar mesons provide the current-component independent predictions in the standard LFQM.

We then apply our current-component independent form factors $(f_\pm, f_T)$ for the 
self-consistent analysis of semileptonic and rare  $D\to (\pi, K)$ decays
using our LFQM constrained by the variational principle for the QCD-motivated effective Hamiltonian with the HO plus
Coulomb interaction~\cite{CJ09,Choi07}.
The form factors $(f_\pm, f_T)$ obtained in the $q^+=0$ frame ($q^2=-{\bf q}^2_\perp <0$) are then analytically continued to the
timelike region by changing ${\bf q}^2_\perp$ to $-q^2$ in the form factors.
In our numerical calculations, we also checked the sensitivity of the constituent charm quark mass $m_c=1.7^{+0.1}_{-0.1}$ GeV
through the analysis of the form factors $(f_\pm, f_T)$ for $D\to (\pi, K)$ decays.
Our results for the form factors and branching ratios for $D\to (\pi, K)$ decays 
show in good agreement with the available experimental data as well as other theoretical predictions. Especially, the smaller charm quark mass $m_c\simeq 1.6$ GeV
seems preferable to larger $m_c=1.8$ GeV for  $D\to (\pi, K)$ decays while they are not much different for the analysis of the decay constant of the $D$ meson.
Finally, we obtained the the $R$ ratios of the semileptonic $D\to (\pi, K)$ decays as a test for the LFU and our
results are consistent with the recent measurements from the BES III~\cite{BES2,BES5} as well as other theoretical results~\cite{FGK20,Ivanov19,Lat18}.

While the rare decay analyses including the tensor form factor can in principle be made, I just focused on the extraction of the current-component 
independent weak transition form factors as well as the comparison with the available experimental data in the present work.  More complete phenomenological 
analyses regarding on the rare decays of heavy $D$ and $B$ mesons are also under consideration.  

\section*{Data Availability}
The data used to support the findings of this study are
available from the  author upon request.

\section*{Conflicts of Interest}
The author declares that there are no conflicts of interest regarding the publication of this paper.

\section*{Acknowledgments}
This work (https://arxiv.org/abs/2108.10544)~\cite{choi21}
 was supported by the National Research Foundation of Korea (NRF) 
under Grant No. NRF- 2020R1F1A1067990.

\end{document}